\documentclass[usenatbib]{mnras}
\usepackage{graphicx}
\usepackage{amsmath}

\usepackage{stix2}

\usepackage{booktabs}

\usepackage{float}

\usepackage{xcolor}

\usepackage{orcidlink}

\usepackage{stfloats}

\title[SAMI and TNG-Cluster]{SAMI and TNG-Cluster: tracing galaxy spin and environmental transformation across cluster phase-space and cosmic time}
\author[G. F. Gon\c{c}alves et al.]{Gustavo F. Gon\c{c}alves$^{1}$\orcidlink{0009-0006-5887-6621}\thanks{E-mail:goncalvesg@alunos.utfpr.edu.br},
Ilaria Marini$^{2}$\orcidlink{0000-0003-2130-2537},
A. Fraser-McKelvie$^{2}$\orcidlink{0000-0001-9557-5648},
Natan de Is\'idio$^{2}$\orcidlink{0009-0005-1424-6604},\newauthor
Rubens E. G. Machado$^{3}$\orcidlink{0000-0001-7319-297X},
Richards P. Albuquerque$^{3}$\orcidlink{0009-0006-9462-1044},
Daudi T. Mazengo$^{2, 4}$\orcidlink{0000-0002-4125-9292}
\newauthor
\\
$^{1}$Departamento Acad\^emico de F\'isica, Universidade Tecnol\'ogica Federal do Paran\'a,
Rua Sete de Setembro 3165, Curitiba, PR, Brazil\\
$^{2}$European Southern Observatory, Karl-Schwarzschild-Strasse 2, 85748 Garching bei M\"unchen, Germany\\
$^{3}$Instituto de Astronomia, Geof\'isica e Ci\^encias Atmosf\'ericas, Universidade de S\~ao Paulo, Rua do Mat\~ao 1226, S\~ao Paulo, SP, Brazil\\
$^{4}$University of Dodoma, College of Natural and Mathematical Sciences, Department of Physics, 1 Benjamin Mkapa Road, 41218,\\ Iyumbu,  Dodoma, Tanzania
}

\date{Accepted XXX. Received YYY; in original form ZZZ}
\pubyear{2026}

\begin{document}  
\label{firstpage}
\pagerange{\pageref{firstpage}--\pageref{lastpage}}
\maketitle

\begin{abstract}
The dense environment of galaxy clusters suppresses star formation and alters the kinematic properties of infalling satellites through gas stripping, tidal interactions, gravitational harassment and starvation. Projected phase-space diagrams connect the present-day distribution of cluster galaxies to their accretion histories.
We combine SAMI Galaxy Survey integral field spectroscopy with the TNG-Cluster simulation to investigate how the stellar spin parameter ($\lambda_R$), $(g-i)$ colour, and sSFR vary across projected phase-space infall regions.
At $z = 0$, TNG-Cluster reproduces the direction and broad strength of the phase-space trends observed in SAMI, including the weak yet significant \(\lambda_R\)–clustercentric distance correlation. Leveraging this agreement, we extend the analysis across the last 8 Gyr, tracing the statistical evolution of galaxy properties within each infall region, and complement this with individual orbital histories of representative satellites.
While colour and sSFR show clear monotonic gradients with both phase-space position and cosmic time, $\lambda_R$ behaves differently: it remains largely uniform across the outer infall regions, with only the virialised core exhibiting systematically lower values, and displays a slow monotonic decline toward the present day across all regions. We find that angular momentum suppression driven by the cluster environment is a slow, cumulative process requiring several Gyr of exposure to the cluster core, modulated by orbital history and stellar mass. This gradual nature explains why the $\lambda_R$--environment correlation appears weak across phase-space regions in statistical samples; the effect emerges when individual satellite histories are tracked, revealing a sustained dynamical response to prolonged cluster residence.
\end{abstract}


\section{Introduction}
Environmental effects in galaxy clusters have long been recognised as an important driver of galaxy evolution. Several physical mechanisms have been proposed to explain how dense environments influence galaxy properties. Observationally, the morphology--density relation shows that early-type galaxies preferentially populate the highest density regions of clusters, while spiral galaxies dominate lower-density environments \citep{Dressler1980, Dressler84}, providing early evidence that the cluster environment drives galaxy transformation. Ram pressure stripping, caused by the interaction between the gaseous disc of a galaxy and the hot intracluster medium, can remove cold gas and affect star formation \citep{Gunn1972}. Repeated high-velocity encounters between galaxies can dynamically heat stellar discs and induce morphological transformations through the process known as galaxy harassment \citep{Moore1996}. In addition, satellites may also be quenched through starvation, whereby the removal of the hot gas reservoir upon infall suppresses future gas cooling and star formation on longer timescales \citep{Larson1980}.

Building on these foundational results, recent studies have extended the observational and theoretical characterisation of these environmental mechanisms across a wide range of environments and galaxy populations. Multi-wavelength observations have revealed examples of ongoing ram pressure stripping in so-called jellyfish galaxies, systems exhibiting extended gaseous tails and disturbed star-forming regions \citep[e.g.,][]{Poggianti2017, Roberts2021, Boselli2022}. More broadly, the role of environment in shaping galaxy properties has been explored through large observational surveys and theoretical models, which operate on timescales of a few Gyr following infall and show that environmental quenching, tidal interactions, and gas stripping contribute jointly to the evolution of satellite galaxies in dense environments \citep[e.g.,][]{Kang2008, Blanton2009, Bahe2017, Rhee2020, Oman2021, Vulcani2021}. Alongside these relatively rapid processes, starvation has emerged as an important long-term quenching pathway, in which the removal of the hot gas reservoir upon infall progressively suppresses future star formation without significantly perturbing the stellar kinematics \citep{Trussler2020, deIsidio2026}.

Numerical simulations complement this observational picture directly. While observations provide statistical views of galaxy populations, simulations allow the evolutionary histories of individual galaxies to be traced across cosmic time, enabling the connection between present-day galaxy properties and their past dynamical evolution within clusters \citep[e.g.,][]{Jaffe2015, LaraLopez22, Goubert25}. In this context, \citet{Jaffe2015} showed that projected phase-space diagrams provide a useful diagnostic to relate the spatial and kinematic distribution of cluster galaxies to their accretion histories and environmental processing.

This diagnostic power arises because different regions of projected phase-space correspond statistically to distinct stages of cluster accretion, as demonstrated by cosmological hydrodynamical simulations. This behaviour is theoretically motivated by the collisionless Boltzmann and Jeans equations, which predict that galaxies at different stages of infall develop systematic gradients in their spatial and kinematic distributions according to their orbital histories, a prediction confirmed by numerical simulations \citep[e.g.,][]{Gill2004, Gill2005, Mahajan2011}. In particular, projected phase-space regions have been associated with variations in time since infall and tidal mass loss \citep[e.g.,][]{Oman2013, Pasquali2019}, providing a framework to interpret the orbital histories of cluster galaxies. This approach allows galaxies to be separated into populations such as recent infallers, backsplash galaxies, and virialised systems using only projected observables, making it applicable both to observations and simulations that are projected to mimic observational measurements \citep{Rhee2017}. 

Projected phase-space diagrams have also proven particularly useful in observational studies to assess cluster membership and constrain the dynamical state of galaxy groups and clusters \citep[e.g.,][]{Biviano2013, Jaffe2015}. As a result, projected phase-space analysis provides a practical framework to investigate how galaxy properties evolve during their infall and subsequent evolution within cluster environments.

IllustrisTNG is particularly well suited to this phase-space approach, as it simultaneously resolves cluster environments and the internal properties of their satellite populations across a broad range of mass scales. The simulation suite includes a set of cosmological volumes with different box sizes and mass resolutions, enabling the study of galaxy formation across a broad range of environments. The simulation model incorporates a comprehensive treatment of baryonic physics, including star formation, stellar feedback, and active galactic nucleus feedback, and has been shown to reproduce several observed galaxy population properties \citep{Springel2018, Pillepich2018a, Nelson2018, Marinacci2018, Naiman2018}.

The suite spans three main volumes: IllustrisTNG50 (hereafter TNG50), IllustrisTNG100 (hereafter TNG100), and IllustrisTNG300 (hereafter TNG300), each offering complementary advantages for environmental studies. TNG50 has been used to investigate the morphological evolution of disc galaxies accreted into cluster environments, finding that most satellites experience structural transformations within $\sim 0.5$--$4$ Gyr after accretion, accompanied by gas removal and star formation quenching \citep{Joshi2020}. TNG100 provides statistically significant cluster samples while maintaining sufficient resolution to study galaxy structure, and has been used to show that repeated pericentric passages lead to substantial dark matter stripping and gradual gas depletion \citep{Lokas2020}. TNG300 extends this to larger statistical samples of massive clusters, enabling studies of star formation triggered by galaxy interactions \citep{Patton2020} and the estimation of galaxy infall times from projected phase-space diagrams \citep{Dou2025}. Across all volumes, IllustrisTNG has been widely used to investigate ram-pressure stripped jellyfish galaxies \citep{Yun2019, Rohr2023, Goller2023, Zinger2024, KurinchiVendhan2025}, demonstrating the capability of cosmological simulations to reproduce a wide range of environmental processes observed in galaxy clusters.

Recently, the TNG-Cluster project \citep{Nelson2024} extended this framework to higher cluster masses through a set of zoom-in simulations performed with numerical resolution comparable to TNG300, spanning a broad range of cluster masses and including a significant number of very massive systems ($M_{200} \gtrsim 10^{15}\,{\rm M_\odot}$)\footnote{$M_{200}$ is defined as the total mass of the host halo enclosed within a sphere of radius $R_{200}$, inside which the mean density is 200 times the critical density of the Universe at the redshift considered. For the simulated sample, this corresponds to the \texttt{Group\_M\_Crit200} field in the TNG catalogues. This definition is adopted throughout the paper.}. This provides an improved statistical sample of massive cluster environments, enabling the identification of simulated analogues of the most massive clusters observed in the local Universe.

On the observational side, recent advances in observational capabilities have enabled new approaches to investigate the internal properties of galaxies in dense environments. In particular, integral field spectroscopy (IFS) provides spatially resolved measurements of galaxy kinematics and stellar populations, allowing the study of structural and dynamical properties beyond what is accessible with single-fibre spectroscopy. Large IFS surveys such as the Mapping Nearby Galaxies at Apache Point Observatory (MaNGA; \citealt{Bundy2015}) and the Sydney-AAO Multi-object Integral field spectrograph (SAMI; \citealt{Croom2012}) have produced statistically significant samples of nearby galaxies with resolved spectroscopic data, enabling systematic studies of galaxy kinematics across a wide range of environments. One of the key dynamical quantities accessible with IFS observations is the spin parameter $\lambda_R$, introduced as a proxy for the projected specific angular momentum of galaxies \citep{Emsellem2007}, which has been widely employed to study morphological changes of galaxies as a function of environment \citep{Fogarty2014, Fogarty2015, Brough2017, Cortese2019, vanDeSande2021, Barsanti2025}. 

The SAMI Galaxy Survey \citep{Croom2012}, which provides spatially resolved spectroscopy for $\sim$3000 nearby galaxies including $\sim$800 satellites across eight galaxy clusters \citep{Owers2017}, represents a particularly powerful test for such studies. Past observational work has established that a kinematic morphology--density relation is indeed in place, with early-type galaxies preferentially occupying the slow-rotator regime in denser environments while disc-dominated systems tend to show higher angular momentum \citep{Brough2017, vanDeSande2021}. However, the direct correlation between $\lambda_R$ and environmental indicators remains weak or statistically insignificant in many cases \citep{vandeSande2017a, Santucci2023}.

At least part of this ambiguity may stem from the fact that observations capture only a single snapshot in time. Cosmological simulations overcome this by granting access to the full evolutionary history of clusters and their satellite populations, directly connecting present-day properties to past environmental processing. In this work, we identify simulated analogues of the observed SAMI clusters within the TNG-Cluster simulations and construct projected phase-space diagrams for both samples to analyse how the spin parameter ($\lambda_{R_e}$), $(g-i)$ colour, and sSFR vary across different accretion regions. To physically motivate these statistical trends, we additionally follow the orbital histories of individual satellite galaxies across all simulation snapshots, tracking their phase-space trajectories alongside properties such as stellar mass growth, gas fraction, star formation rate, $(g-i)$ colour, and $\lambda_{R_e}$.

\begin{table*}
\centering
\caption{Properties of the eight SAMI clusters and the number of galaxies available for the phase-space analysis. The cluster properties ($z_{\rm cl}$, $M_{200}$ and $R_{200}$) are taken from \citet{Owers2017}. Galaxy identifiers follow the SAMI Galaxy Survey catalogue described by \citet{Bryant2015}. Stellar kinematic measurements, including $\lambda_R$, are taken from the SAMI stellar kinematics catalogue of \citet{vandeSande2017a}. $N_{\rm sat}$ is the total number of satellite galaxies with $R_{\rm proj}/R_{200}\leq 3$ available in each cluster. The columns $N_{\lambda_R}$, $N_{(g-i)}$, and $N_{\log(\mathrm{sSFR})}$ give the number of satellite galaxies with phase-space coordinates for which the corresponding property is available (and thus used to define the colormap of the phase-space panels).}
\label{tab:sami_clusters_phase_space}
\begin{tabular}{lccccccccc}
\toprule
Cluster & $z_{\rm cl}$ & $\log(M_{200}/{\mathrm M_\odot})$ & $R_{200}$ (Mpc) &
$N_{\rm sat}$ &
$N_{\lambda_R}$ &
$N_{(g-i)}$ &
$N_{\mathrm{sSFR}}$ \\
\midrule
Abell119   & 0.0442 & 14.92 & 2.02 & 180 & 140 & 180 & 71  \\
Abell85    & 0.0549 & 15.19 & 2.42 & 167 & 114 & 167 & 91  \\
Abell2399  & 0.0579 & 14.66 & 1.63 & 137 & 97  & 137 & 78  \\
Abell4038  & 0.0293 & 14.36 & 1.46 & 97  & 82  & 97  &         0 \\
Abell3880  & 0.0578 & 14.64 & 1.62 & 117 &        69 & 117 &         0 \\
Abell168   & 0.0449 & 14.28 & 1.33 & 84  & 60  & 84  & 56  \\
APMCC0917  & 0.0509 & 14.26 & 1.19 & 41  & 25  & 41  &         0 \\
EDCC0442   & 0.0498 & 14.45 & 1.41 & 45  & 34  & 45  &         0 \\
\midrule
Total &  &  &  & 868 & 621 & 868 & 296 \\
\bottomrule
\end{tabular}

\begin{minipage}{0.98\textwidth}
\footnotesize
\end{minipage}
\end{table*}

\section{Observational sample : SAMI Clusters}
\label{obs}

The Sydney-AAO Multi-object Integral field spectrograph (SAMI) Galaxy Survey was an integral field spectroscopy survey that observed 3068 galaxies with the Anglo-Australian Telescope between 2013 and 2018 \citep{Croom2012, Bryant2015}. A unique feature of the SAMI survey was its dedicated cluster programme \citep{Owers2017}, targeting $\sim$800 galaxies located in eight nearby galaxy clusters, making it one of the largest IFS samples of cluster galaxies available to date.

Our observational sample consists of eight galaxy clusters observed as part of the SAMI Galaxy Survey. The cluster sample spans the redshift range $0.03 < z < 0.06$ and covers halo masses between $\log(M_{200}/{\mathrm M_\odot}) \sim 14.3$ and $15.2$. The cluster properties, including redshift, halo virial mass ($M_{200}$), and virial radius ($R_{200}$), are taken from \citet{Owers2017}. Satellite galaxies are identified using the spectroscopic membership defined in the SAMI cluster catalogues. To ensure consistency with the simulated sample introduced in Section~\ref{sim}, we restrict the analysis to satellite galaxies with stellar masses $\log(M_\star/{\rm M_\odot}) \geq 9.75$. The spectroscopic completeness within $R < 2R_{200}$ reaches $\gtrsim 91$ per cent for most clusters in the sample, with a median completeness of $\sim 96$ per cent, ensuring that the satellite population is well sampled across the full radial range considered \citep{Owers2017}. For the phase-space analysis we consider galaxies within a projected radius of $R_{\rm proj}/R_{200} \leq 3$ relative to the cluster centre.

Table~\ref{tab:sami_clusters_phase_space} summarises the main properties of the cluster sample and the number of satellite galaxies available for the phase-space analysis in each system.

\subsection{SAMI clusters and galaxies properties}

For each satellite galaxy we analyse three properties that trace different aspects of galaxy evolution: the stellar spin parameter $\lambda_R$, the optical colour $(g-i)$, and the sSFR. Stellar kinematic measurements, including the spin parameter $\lambda_R$, are taken from the SAMI stellar kinematics catalogue of \citet{vandeSande2017a}. Galaxy identifiers and photometric quantities follow the SAMI Galaxy Survey catalogues described by \citet{Bryant2015}. 

Optical colours $(g-i)$ are obtained from the photometric catalogues associated with the SAMI survey \citep{Bryant2015}, while sSFR are derived from emission-line measurements \citep{Medling2018}. The availability of each property varies among clusters due to differences in observational coverage and data quality. Note that the number of galaxies with measurements varies between quantities, as detailed in Table~\ref{tab:sami_clusters_phase_space}.

\begin{figure*}
\centering
\includegraphics{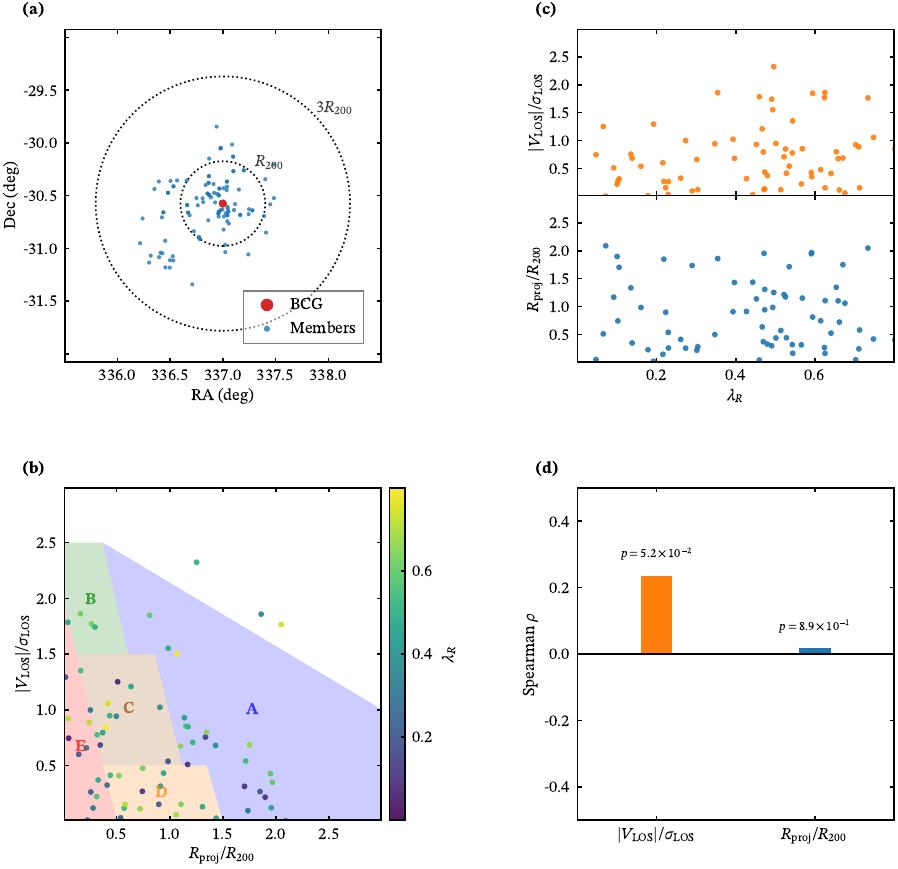}
\caption{Example of the methodology adopted to build the phase-space analysis (illustrated for the cluster Abell~3880).
~\textbf{(a)} Sky distribution of the BCG and satellite members in the RA--Dec plane. The projected radii $R_{200}$ and $3R_{200}$ are indicated.
~\textbf{(b)} Projected phase-space diagram of satellite galaxies with available $\lambda_R$ measurements, shown as $V_{\rm LOS}/\sigma_{\rm LOS}$ versus $R_{\rm proj}/R_{200}$, where $V_{\rm LOS}$ is the line-of-sight velocity relative to the BCG and $\sigma_{\rm LOS}$ is the cluster velocity dispersion estimated from satellites within $R_{200}$. The background regions (A--E) mark the infall classes defined by \citet{Rhee2017}, ranging from early infallers (region~A) to ancient infallers (region~E). Galaxies are colour-coded by their $\lambda_R$.
~\textbf{(c)} Auxiliary relations used to quantify trends in panel~(b): (top) $V_{\rm LOS}/\sigma_{\rm LOS}$ as a function of $\lambda_R$; (bottom) $R_{\rm proj}/R_{200}$ as a function of $\lambda_R$.
~\textbf{(d)} Spearman rank correlation coefficients for the two relations shown in panel~c). Bar heights indicate the Spearman $\rho$ values, and the corresponding $p$-values are reported above each bar.}
\label{fig:fig1}
\end{figure*}

To investigate environmental trends within clusters we construct projected phase-space diagrams using the clustercentric distance and the line-of-sight velocity of satellite galaxies. The construction of the diagram follows the steps illustrated in Fig.~\ref{fig:fig1}. Panel~(a) shows the sky distribution of the cluster members in the RA--Dec plane, where the brightest cluster galaxy (BCG) defines the cluster centre and the projected radii $R_{200}$ and $3R_{200}$ are indicated, with $R_{200}$ taken from \citet{Owers2017}. In panel~(b), the projected phase-space diagram is built by combining the projected clustercentric distance $R_{\rm proj}/R_{200}$ with the normalised line-of-sight velocity $|V_{\rm LOS}|/\sigma_{\rm LOS}$, where $V_{\rm LOS}$ is measured relative to the BCG redshift and $\sigma_{\rm LOS}$ corresponds to the cluster velocity dispersion estimated from satellites within $R_{200}$, both derived from the spectroscopic catalogues of \citet{Owers2017}. The background regions (A--E) follow the infall class boundaries defined by \citet{Rhee2017}, who used cosmological hydrodynamical simulations to associate distinct regions of projected phase-space with different accretion epochs. Region A corresponds to the most recently accreted galaxies, still located at large clustercentric radii and high line-of-sight velocities, while region E represents the ancient, virialised population concentrated in the cluster core. Regions B, C, and D trace intermediate stages of infall, providing a statistical framework to interpret the orbital histories of satellite galaxies from projected observables alone. In panel~(c), the relations used to quantify possible trends are shown, namely $|V_{\rm LOS}|/\sigma_{\rm LOS}$ and $R_{\rm proj}/R_{200}$ as a function of the spin parameter $\lambda_{R_e}$. Finally, panel~(d) summarises these relations through the Spearman rank correlation coefficients, computed between each galaxy property and the two phase-space coordinates across the full stacked sample. The bar heights indicate the corresponding $\rho$ values, which measure the monotonicity of the relation, and the associated $p$-values reported above each bar indicate the probability of obtaining such a correlation by chance under the null hypothesis of no correlation.

\begin{figure*}
\centering
\includegraphics{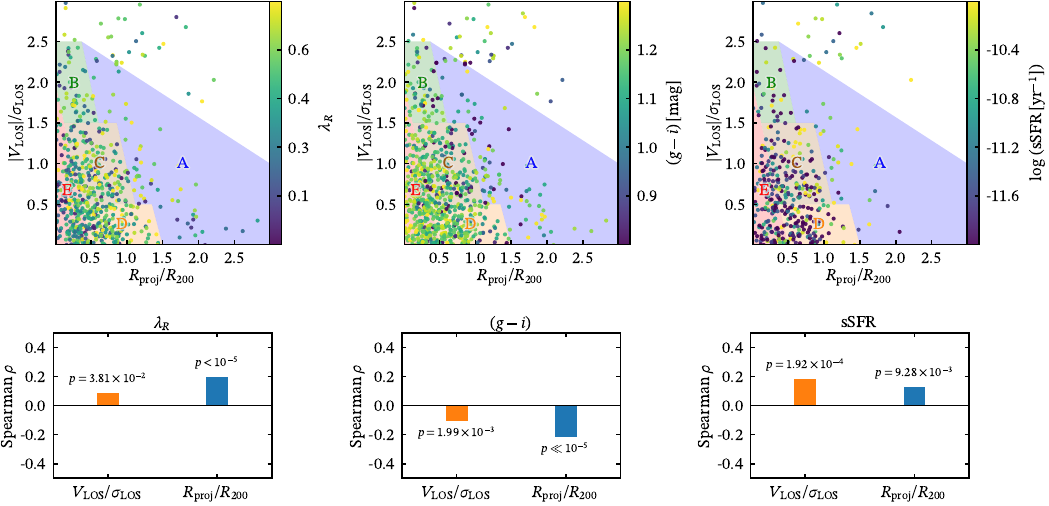}
\caption{Stacked projected phase-space diagrams for the eight SAMI clusters. From left to right, satellite galaxies are mapped to the colour scale by $\lambda_R$, $(g-i)$, and $\log(\mathrm{sSFR})$. Each top panel shows $V_{\rm LOS}/\sigma_{\rm LOS}$ as a function of $R_{\rm proj}/R_{200}$ for the stacked cluster sample, considering only galaxies with the corresponding property available. The panels below report the Spearman rank correlation coefficients ($\rho$) for the relations between each property and the two phase-space coordinates, with the associated $p$-values annotated next to the bars.}
\label{fig:fig2}
\end{figure*}

Because the number of galaxies per individual cluster is relatively small, especially for galaxies with available kinematic measurements, the subsequent analysis is performed using stacked phase-space diagrams combining all clusters in the sample, as shown in Fig.~\ref{fig:fig2}. Particular attention should be given to the reported $p$-values, as they provide a quantitative measure of the statistical significance of the trends; throughout this work, correlations with $p < 0.05$ are considered statistically significant. For some properties, the correlations do not reach this significance level, which may reflect either a genuine absence of environmental dependence or the limited statistical power of the observational sample. These properties will serve as a reference for the analogous analysis performed later with the larger simulation sample.

\section{Simulated Sample: TNG-Cluster}
\label{sim}

The IllustrisTNG project is a cosmological hydrodynamical simulation suite performed with the moving-mesh code AREPO \citep{Springel2010}. The galaxy formation and evolution model adopted in these simulations is described in detail in \citet{Weinberger2017} and \citet{Pillepich2018a}. The suite consists of several simulation runs that progressively increase the cosmological volume while correspondingly reducing the numerical resolution. The simulations adopt the cosmological parameters from the Planck Collaboration \citep{Planck2016}. These correspond to a matter density $\Omega_{\rm m} = \Omega_{\rm dm} + \Omega_{\rm b} = 0.3089$, a baryon density $\Omega_{\rm b} = 0.0486$, a cosmological constant $\Omega_\Lambda = 0.6911$, and a Hubble constant $H_0 = 100\,h\,\mathrm{km\,s^{-1}\,Mpc^{-1}}$ with $h = 0.6774$. The normalization of the matter power spectrum is $\sigma_8 = 0.8159$, and the spectral index of the primordial power spectrum is $n_s = 0.9667$. The initial conditions of the simulations were generated at redshift $z = 127$.

TNG-Cluster \citep{Nelson2024} consists of 352 zoom-in simulations of galaxy clusters and adopts the same numerical resolution as TNG300, with dark matter particle masses $m_{\rm DM} \approx 5.9\times10^7\,\mathrm{M_\odot}$ and mean gas cell masses $m_{\rm gas} \approx 1.1\times10^7\,\mathrm{M_\odot}$. Its large sample of massive systems provides the statistical coverage required to identify simulated analogues of the SAMI clusters considered in this work. At the low-mass end of the SAMI cluster sample, where TNG-Cluster provides fewer suitable analogues, we supplement the sample with halos from TNG300, following \citet{Nelson2024}, who describe the two simulations as complementary. Our goal is to identify simulated cluster environments that statistically represent the mass range of the eight clusters in the SAMI sample and to analyse the properties of satellite galaxies in different cluster regions using projected phase-space diagrams. Some of these quantities, such as integrated colors, are relatively insensitive to the spatial resolution of the simulation. Others depend more directly on the numerical resolution, particularly the stellar spin parameter $\lambda_R$, which depends on the number of stellar particles used to resolve each galaxy. The definition of the spin parameter is discussed in more detail in Section~\ref{spin parameter simulation}.

\subsection{Cluster sample selection and satellite galaxies}
\label{cluster sample selection}

\begin{table*}
\centering
\caption{Observed SAMI clusters and their corresponding representation in the TNG-Cluster sample. 
For each observed cluster mass, we select simulated halos with similar $M_{200}$. 
The table lists the SAMI cluster name, the target cluster halo mass, the median mass of the selected TNG halos, 
and the total number of satellite galaxies across the selected halos.}
\label{tab:sami_representation}
\begin{tabular}{lcccc}
\toprule
SAMI Cluster & $\log(M_{200}/{\mathrm M_\odot})$ & ${\rm TNG \ }N_{\rm cluster}$ & ${\rm TNG \ }\log(M_{200}^{\rm med}/{\mathrm M_\odot)}$ & ${\rm TNG \ }N_{\rm sats}$ \\
\midrule
APMCC0917$^\dagger$  & 14.26 & 30 & 14.26 & 1182 \\
Abell 168$^\dagger$  & 14.28 & 26  & 14.27 & 1137 \\
Abell 4038  & 14.36 & 22 & 14.41 & 1177 \\
EDCC0442   & 14.45 & 25 & 14.49 & 1511 \\
Abell 3880  & 14.64 & 44 & 14.64 & 3809 \\
Abell 2399  & 14.67 & 29 & 14.73 & 2949 \\
Abell 119   & 14.93 & 44 & 14.93 & 6992 \\
Abell 85    & 15.19 & 41 & 15.16 & 11592 \\
\midrule
Total &  & 261 &  & 30349 \\
\bottomrule
\multicolumn{5}{l}{$^\dagger$ Supplemented with halos from TNG300 to account for the limited number of analogues} \\
\multicolumn{5}{l}{\phantom{$^\dagger$} at this cluster mass range in TNG-Cluster.} \\
\end{tabular}
\end{table*}

To construct a simulated cluster sample comparable to the observed SAMI clusters, we identify halos at $z=0$ in the TNG-Cluster simulation whose virial masses lie within the mass range spanned by the observed systems. The SAMI sample contains eight clusters with halo masses $M_{200}$ between $\log(M_{200}/{\mathrm M_\odot})~\sim~14.3$ and $15.2$. We therefore select halos from the full TNG-Cluster catalogue whose masses fall within a tolerance window around these values. Specifically, for each SAMI cluster mass we identify simulated halos within a fractional range of $\pm20$ per cent and randomly sample from this pool, ensuring that halos are not reused. This procedure yields a simulated sample of 221 clusters whose mass distribution broadly follows that of the SAMI systems, although the two lowest-mass clusters (APMCC0917 and Abell~168, $\log(M_{200}/{\rm M_\odot}) \lesssim 14.3$) required supplementation with halos from TNG300 to achieve comparable statistical representation.

Satellite galaxies are defined as subhalos gravitationally bound to a given host halo, excluding the central galaxy. In the TNG data structure these correspond to subhalos identified by the SUBFIND algorithm \citep{Springel2001, Dolag2009} that belong to the same Friends-of-Friends (FoF) group as the host cluster. To ensure reliable measurements of galaxy kinematics, we restrict our analysis to satellite galaxies with stellar masses $M_\star \geq 10^{9.75}~{\mathrm M_\odot}$. Given the stellar mass resolution of the TNG-Cluster simulation, this limit corresponds to systems containing at least $\sim 500$ stellar particles.

Table~\ref{tab:sami_representation} summarizes the correspondence between the observed clusters and their representation in the TNG-Cluster dataset. For each SAMI cluster, the table lists the target halo mass,  the number of matched TNG-Cluster systems, the median mass of the selected simulated halos, and the total number of satellite galaxies identified across the selected systems.

Although other quantities considered in this work, such as $(g-i)$ colour, are less sensitive to the number of stellar particles and could in principle be measured for lower-mass systems, we apply the same stellar mass limit to all properties. This ensures that the phase-space diagrams presented for different galaxy properties correspond to the same underlying galaxy sample, allowing for a consistent comparison between panels.

This selection results in a simulated sample approximately 1.5 orders of magnitude larger in the number of clusters and in the number of satellite galaxies compared to the observational sample (Table~\ref{tab:sami_representation}), reflecting the fact that the simulation provides complete information for all resolved galaxies, while the observations are subject to the usual selection and completeness limitations. To ensure a consistent comparison between the two samples, the same stellar mass threshold of $\log(M_\star/{\rm M_\odot}) \geq 9.75$ is applied to both the observational and simulated satellite populations. This difference is particularly relevant at projected clustercentric distances beyond $R_{\rm proj}/R_{200} > 1$, where the SAMI observations contain relatively few galaxies, while the simulation provides a substantially larger number of satellite systems.

In the simulated version of the projected phase-space analysis presented in this work, we explore three physical properties of satellite galaxies using color-coded phase-space diagrams. These include sSFR, $\lambda_{R}$ and the integrated colour $(g-i)$. The SFR is quantified through \texttt{SubhaloSFR} field, which corresponds to the sum of the star formation rates of all gas cells bound to the subhalo, normalized by the stellar mass. In IllustrisTNG, galaxies with no actively star-forming gas cells are assigned $\mathrm{SFR} = 0$ by default, as the simulation does not impose a detection threshold analogous to observational limits \citep{Donnari2019, Davies2019}. These systems are commonly identified as quenched galaxies in the simulation context. For such galaxies, we assign a floor value of $\log(\mathrm{sSFR}) = -12$ in order to include them in the logarithmic representation. Integrated galaxy colors are computed from the \texttt{SubhaloStellarPhotometrics} magnitudes provided in the $g$ and $i$ bands.

In addition to these quantities, we also analyze the stellar spin parameter $\lambda_R$, which is not directly provided in the simulation catalogs and must be computed from the stellar kinematics of each galaxy. The methodology adopted to derive $\lambda_R$ is described in detail in Section~\ref{spin parameter simulation}.

\subsection{Simulation spin parameter definition}
\label{spin parameter simulation}
To quantify the stellar dynamical support of galaxies in the simulations we adopt the dimensionless spin parameter $\lambda_R$. Our methodology follows the formalism introduced by \citet{Lagos2017} and later applied to TNG100 by \citet{Pallero2025}. In our implementation, stellar particles are recentred on the centre-of-mass of each subhalo and projected onto the galaxy plane. The half-mass radius is computed from the stellar particle distribution and used to define the radial aperture of the analysis. The $\lambda_R$ parameter is then evaluated within a set of cumulative radial bins of fixed width, where the rotational velocity is estimated from the mass-weighted specific angular momentum of the particles and the velocity dispersion from the component of the stellar velocities parallel to the total angular momentum vector.

We first compute the stellar specific angular momentum vector

\begin{equation}
\mathbf{j}_\star =
\frac{\sum_i m_i (\mathbf{r}_i - \mathbf{r}_{\mathrm{COM}}) \times (\mathbf{v}_i - \mathbf{v}_{\mathrm{COM}})}
{\sum_i m_i},
\end{equation}

\noindent where $m_i$ is the mass of the $i$-th stellar particle, $\mathbf{r}_i$ and $\mathbf{v}_i$
are its position and velocity vectors, and $\mathbf{r}_{\mathrm{COM}}$ and
$\mathbf{v}_{\mathrm{COM}}$ correspond to the centre-of-mass position and velocity
of the galaxy. The direction of $\mathbf{j}_\star$ defines the stellar angular momentum axis
$\mathbf{L}_\star$ used as the reference axis for the kinematic decomposition.
The velocity of each particle relative to the centre-of-mass is

\begin{equation}
\Delta \mathbf{v}_i = \mathbf{v}_i - \mathbf{v}_{\mathrm{COM}} .
\end{equation}

\noindent The component of the velocity parallel to the angular momentum axis is

\begin{equation}
v_{\parallel,i} =
\frac{\Delta \mathbf{v}_i \cdot \mathbf{L}_\star}{|\mathbf{L}_\star|}.
\end{equation}

\noindent Using this projection we compute the one-dimensional velocity dispersion

\begin{equation}
\sigma_{1D} =
\sqrt{
\frac{\sum_i m_i v_{\parallel,i}^2}
{\sum_i m_i}
}.
\end{equation}

\noindent The rotational velocity at radius $r_k$ is defined as

\begin{equation}
V_{\mathrm{rot}}(r_k) =
\frac{|\mathbf{j}_\star(r < r_k)|}{r_k},
\end{equation}

\noindent where $\mathbf{j}_\star(r < r_k)$ is the specific angular momentum computed using
all stellar particles enclosed within the cumulative aperture $r_k$. Finally, the stellar spin parameter $\lambda_R$ is defined as \citep{Emsellem2007}

\begin{equation}
\lambda_R =
\frac{\sum_i m_i r_i V_{\mathrm{rot},i}}
{\sum_i m_i r_i \sqrt{V_{\mathrm{rot},i}^2 + \sigma_{1D,i}^2}} .
\end{equation}

\noindent All quantities are calculated using stellar particles within three times the galaxy's stellar half-mass radius.

\subsection{Resolution limits and the interpretation of $\lambda_R$}
\label{sec:caveats}

The finite mass resolution of TNG-Cluster and TNG300 limits the internal stellar kinematics that can be recovered for individual galaxies, particularly near our stellar-mass threshold of $\log(M_\star/\mathrm{M_\odot})=9.75$, where satellites contain only
$\sim500$ stellar particles. We therefore do not interpret the simulated $\lambda_R$ measurements as resolved mock-IFS measurements. Instead, they are used as a particle-based proxy for the degree of stellar rotational support.

The aperture of $3\,R_{1/2,\star}$ adopted in Section~\ref{spin parameter simulation}
follows \citet{Lagos2017} and \citet{Pallero2025}. Restricting the measurement to a smaller aperture would substantially reduce the number of stellar particles contributing to the kinematic estimate, especially for galaxies close to the mass limit. The larger aperture therefore provides a better-sampled estimate of the rotational structure resolved
by the simulation. We stress, however, that this aperture is not equivalent to the projected effective-radius aperture used for the SAMI measurements. Environmental changes occurring preferentially at large galactocentric radii may consequently affect the simulated and observed measurements differently.

For this reason, our comparison with SAMI is restricted to the relative behaviour of $\lambda_R$ with environment and time, rather than its absolute normalisation. Accordingly, the simulated and observed \(\lambda_R\) measurements should not be interpreted on a galaxy-by-galaxy basis or as measurements on the same absolute scale. Appendix~\ref{appendix C} examines the reliability of this interpretation in detail. The stellar-particle maps show that high- and low-spin systems retain clearly different kinematic structures across the adopted stellar-mass range, including close to the
mass threshold (Fig.~\ref{fig:figC1}), while the $\lambda_R$--$\varepsilon$ relation remains present in the lowest-mass galaxies (Fig.~\ref{fig:figC2}). Figures~\ref{fig:figC3} and \ref{fig:figC4} further show how the measured $\lambda_R$ values relate to the resolved stellar velocity structure across the simulated sample.

\subsection{Simulated projected phase-space diagrams}
\label{phase-space}

Projected phase-space diagrams for the simulated clusters are constructed following the same definitions adopted for the observational analysis. For each halo in our TNG-Cluster sample we compute the projected clustercentric distance and the velocity offset of satellite galaxies relative to the cluster center.

The cluster center is defined as the position of the central galaxy of the corresponding FoF halo, identified through the \texttt{GroupFirstSub} field in the TNG catalogs. Satellite galaxies correspond to subhalos belonging to the same FoF group, excluding the central object. Positions and velocities are taken from the subhalo catalogs and converted to physical units.

To mimic the observational phase-space construction, galaxy positions are projected onto the $xy$ plane of the simulation box to obtain the projected clustercentric distance, which is normalized by the halo radius $R_{200}$ to define $R_{\rm proj}/R_{200}$. Velocities are measured relative to the central galaxy, and the line-of-sight component is taken along the $z$ axis of the simulation box. The velocity coordinate of the phase-space diagram is defined as the absolute line-of-sight velocity normalized by the cluster velocity dispersion, $|V_{\rm LOS}|/\sigma_{\rm LOS}$, where $\sigma_{\rm LOS}$ is the line-of-sight velocity dispersion of satellite galaxies within $R_{200}$.

For the phase-space analysis we consider galaxies within a maximum three-dimensional distance of $3\,R_{200}$ from the cluster center, ensuring that both virialized and infall regions are sampled. The resulting projected phase-space coordinates are then combined with the galaxy properties described in the previous section to produce the color-coded phase-space diagrams used throughout this work.

\subsection{Tracking cluster systems across simulation snapshots}
\label{tracking}

The cluster sample is defined at $z=0$ according to the selection criteria described in Section~\ref{cluster sample selection}. In order to study their evolution, we identify the corresponding systems at earlier epochs using the simulation snapshots 99, 84, 67, and 50, which correspond to redshifts of $z = 0.0$, $0.2$, $0.5$, and $1.0$, respectively.

Tracking the merger trees of all satellite galaxies in order to identify their progenitors at earlier snapshots is, in practice, a challenging task for large samples such as ours. Performing this procedure over the full satellite population and across multiple snapshots is also computationally demanding. In addition, the identification of satellite progenitors can be ambiguous during the infall process and during strong interactions, particularly at the mass resolution of the TNG-Cluster simulation. These effects may introduce discontinuities or misidentifications in the merger trees of low-mass satellites.

Instead, we adopt a cluster-centric tracking strategy based on the brightest cluster galaxy (BCG). The BCGs are massive systems with well-defined merger trees that can be robustly traced across snapshots. For each cluster selected at $z=0$, we identify its central galaxy and follow the main progenitor branch of this subhalo using the SubLink merger tree. At each target snapshot, we determine the FoF halo in which the BCG progenitor resides and identify all subhalos belonging to that FoF group. These objects are considered the cluster members at that epoch, from which the satellite population is selected according to our stellar mass limit. At each target snapshot, the projected phase-space diagram is reconstructed independently using the $R_{200}$ and $\sigma_{\rm LOS}$ of the corresponding host halo at that epoch.

Galaxy properties are studied statistically within regions of the projected phase-space diagram, allowing us to investigate how $\lambda_R$, $(g-i)$ colour, and sSFR vary across different dynamical environments within the cluster. To physically interpret the trends observed across these regions, we additionally trace the full orbital and physical histories of a subset of individual satellite galaxies across simulation snapshots, following their trajectories through the cluster potential alongside properties such as stellar mass growth, gas fraction, star formation rate, $(g-i)$ colour, and $\lambda_R$.

Additional information about the sample distribution is provided in Appendix~\ref{appendix A}. In particular, Fig.~\ref{fig:figA1} shows the stellar mass distribution of the galaxies within each infall region at the different redshifts considered in this work. Furthermore, Table~\ref{tab:region_counts} lists the corresponding galaxy counts and fractions in each region, providing a quantitative overview of the sample composition.

\subsection{Individual satellite orbital histories}
\label{individual satellite}

To complement the statistical analysis across phase-space regions, we trace the full orbital and physical histories of a sample of representative satellite galaxies selected from a single cluster in the TNG300 simulation at $z = 0$. The choice of TNG300 rather than TNG-Cluster at this stage is motivated by resolution considerations: satellites at large clustercentric distances, prior to their first infall, would otherwise be outside the TNG300-resolution zoom-in region of TNG-Cluster, potentially introducing numerical artefacts in their measured properties.

The satellite galaxies selected for this analysis were chosen to illustrate a range of distinct infall histories, and were identified by requiring that their progenitors could be robustly traced as members of the same FoF halo across multiple simulation snapshots. In practice, not all satellite galaxies satisfy this requirement, as merger tree discontinuities and ambiguous group membership during infall can interrupt the progenitor chain. The galaxies presented here therefore represent cases where a complete and unambiguous orbital history could be recovered.

For each of these galaxies, we follow the main progenitor branch using the SubLink merger tree, adopting the same cluster-centric tracking strategy described in Section~\ref{tracking}. At each snapshot, we verify that the progenitor belongs to the FoF halo identified via the BCG main progenitor branch, and extract its physical properties, including stellar mass, gas fraction, SFR, $(g-i)$ colour, and $\lambda_R$. The projected phase-space position of each satellite at each epoch is computed by constructing the full phase-space diagram of the host cluster at that snapshot, following the procedure described in Section~\ref{phase-space}. The epoch of first $R_{200}$ crossing is identified as the snapshot at which the three-dimensional distance between the satellite progenitor and the cluster centre first falls below $R_{200}$. We note that, due to projection effects, the projected phase-space position of a satellite does not always reflect its true three-dimensional location within the cluster, and this should be kept in mind when interpreting the orbital trajectories.

\section{Results and Discussion}
\label{results}

\begin{figure*}
\centering
\includegraphics{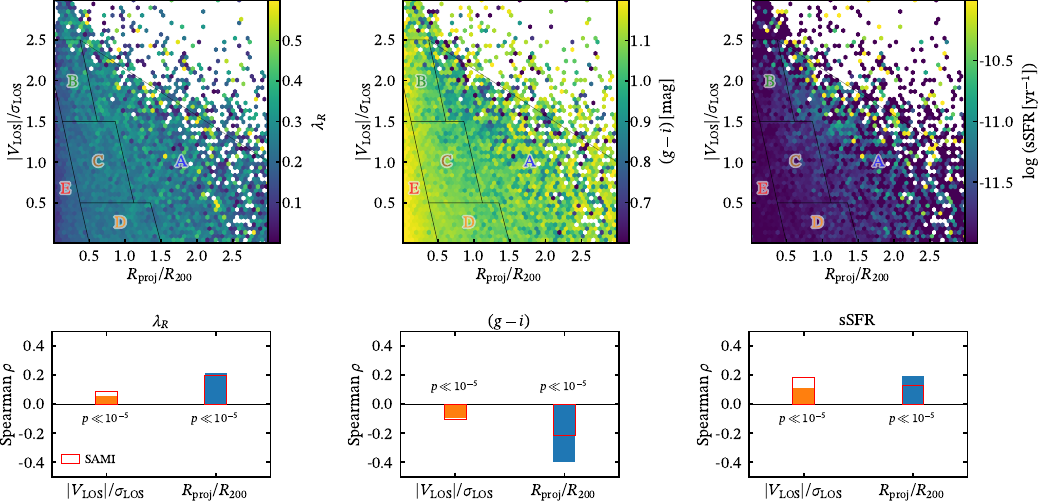}
\caption{Same as Fig.~\ref{fig:fig2}, but for the TNG-Cluster sample. Owing to the much larger number of simulated clusters and satellite galaxies, the stacked phase-space distributions are shown using hexagonal binning to visualise the point density. From left to right, the colour scale encodes $\lambda_R$, $(g-i)$, and $\log(\mathrm{sSFR})$. The panels below report the Spearman rank correlation coefficients ($\rho$) between each property and the two phase-space coordinates, with the corresponding $p$-values indicated.}
\label{fig:fig3}
\end{figure*}

\begin{figure*}
\centering
\includegraphics{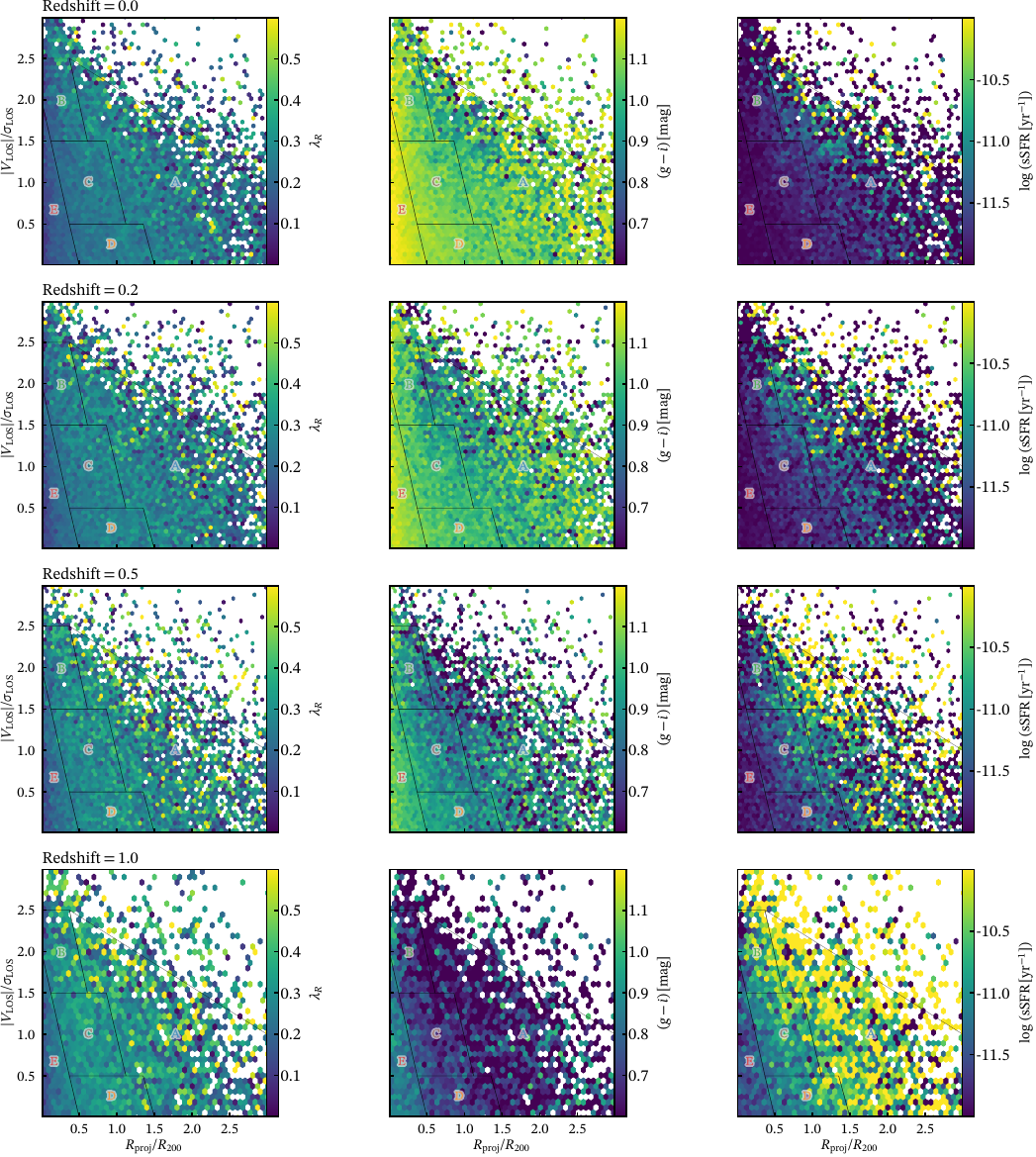}
\caption{Stacked projected phase-space diagrams for the TNG-Cluster sample, shown at four simulation snapshots following the same cluster evolutionary history traced from the $z=0$ BCG. Each row corresponds to a redshift of $z = 0.0$, $0.2$, $0.5$, and $1.0$ (from top to bottom). From left to right, the three columns map the phase-space distribution using the same set of properties as in Fig.~\ref{fig:fig3}: $\lambda_R$, $(g-i)$, and $\log(\mathrm{sSFR})$. Owing to the large number of simulated satellites, the phase-space distributions are visualised with hexagonal binning; the hexbin grid resolution is adjusted between epochs to account for the changing number of available galaxies.}
\label{fig:fig4}
\end{figure*}

\begin{figure*}
\centering
\includegraphics[]{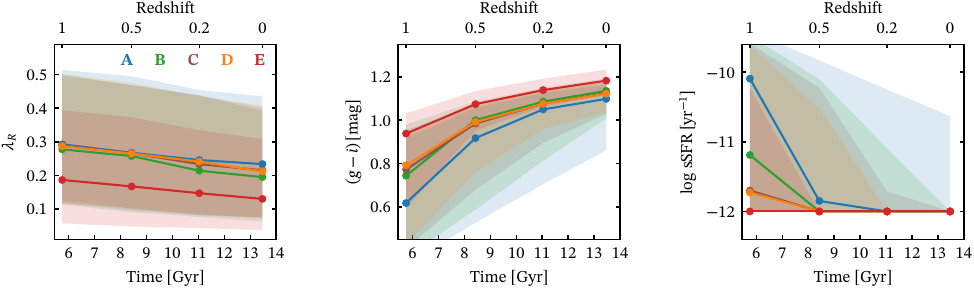}
\caption{Evolution of median galaxy properties across projected phase-space infall regions for the combined TNG-Cluster sample. From left to right, panels show the median values of $\lambda_R$, $(g-i)$, and $\log(\mathrm{sSFR})$ as a function of cosmic time for regions A--E (following the infall-class boundaries defined by \citealt{Rhee2017}). Coloured lines and points indicate the regional medians, and shaded bands show the 16th--84th percentile range of the distribution in each region, using the same colour scheme adopted throughout this work.}
\label{fig:fig5}
\end{figure*}

\subsection{Matching SAMI and TNG-Cluster at z = 0}
In order to compare the environmental trends observed in the SAMI sample with those predicted by the simulation, in Fig.~\ref{fig:fig3} we analyse the distribution of three galaxy properties in projected phase-space diagrams: the spin parameter $\lambda_R$, $(g-i)$ colour, and sSFR. For each property we compute the Spearman rank correlation coefficient ($\rho$) with the two phase-space axes, $R_{\rm proj}/R_{200}$ and $|V_{\rm los}|/\sigma_{\rm los}$.

It is important to stress that this analysis does not aim to compare the absolute values of the galaxy properties between the observations and the simulation. Instead, our goal is to quantify how these properties vary as a function of clustercentric distance and line-of-sight velocity. This approach allows us to assess whether the environmental trends predicted by the simulation match those observed in SAMI, without requiring the
absolute normalisation of the measured quantities to be identical between
the two datasets.

An additional caveat specific to $\lambda_{R_e}$ concerns the different aperture definitions adopted in observations and simulations: in the SAMI data, $\lambda_{R_e}$ is measured within one projected half-light radius, whereas in the simulations we compute $\lambda_R$ within three times the stellar half-mass radius (Section~\ref{spin parameter simulation}). The larger aperture in the simulations is adopted to maximise the number of stellar particles contributing to the kinematic estimate, which is particularly important for lower-mass galaxies where the finite numerical resolution limits the available particle counts. Since these apertures are not equivalent, the absolute values of $\lambda_R$ are not directly comparable between the two samples, and this should be kept in mind when interpreting the results. The agreement discussed below therefore refers to the direction and relative strength of the population-level phase-space correlations, rather than to the absolute \(\lambda_R\) values of individual galaxies.

Overall in Fig.~\ref{fig:fig3}, all three properties show Spearman coefficients in the same direction in both SAMI and TNG-Cluster, and the values of $\rho$ are generally similar. This indicates that the simulation reproduces the main environmental trends observed in the data. One important difference is the statistical significance of the correlations. In the SAMI sample, the limited number of satellites, particularly at radii beyond $R_{200}$, results in low statistical significance for some relations. In contrast, the much larger number of galaxies available in TNG-Cluster leads to statistically significant correlations in most cases. This suggests that some of the weak trends seen in the observations were likely real but limited by sample size.

In the first leftmost panel of Fig.~\ref{fig:fig3}, spin parameter $\lambda_R$ shows the closest agreement between observations and simulation. The Spearman coefficients are nearly identical in both datasets. A weak but statistically significant correlation is present with $|V_{\rm los}|/\sigma_{\rm los}$ ($\rho \sim 0.1$), while a higher correlation coefficient is found with clustercentric distance ($\rho \sim 0.19$). This indicates that galaxies located in the inner regions of clusters tend to have lower spin parameters, consistent with the expectation that dense environments promote dynamical processes capable of reducing galaxy angular momentum \citep[e.g.,][]{Fogarty2014, Fogarty2015, Brough2017, Cortese2019, vanDeSande2021, Santucci2023, Barsanti2025}.

In the second panel of Fig.~\ref{fig:fig3}, both the observations and the simulation show a clear abundance gradient in colour. Blue galaxies are preferentially located in the outer regions of the cluster, producing a clear negative correlation with clustercentric distance. The Spearman coefficient with $R_{\rm proj}/R_{200}$ is $\rho \sim -0.20$ in SAMI and $\rho \sim -0.40$ in TNG-Cluster. The stronger trend in the simulation may be related to differences in radial coverage between the two samples. In particular, the TNG-Cluster sample includes galaxies out to larger radii, where the fraction of blue galaxies increases significantly, while most SAMI satellites lie within $R_{200}$, with fewer objects at larger distances. A weaker ($\rho \sim -0.1$) but still significant correlation is also present with $|V_{\rm los}|/\sigma_{\rm los}$.

In the third panel of Fig.~\ref{fig:fig3}, the behaviour of the sSFR is also broadly consistent between the two datasets. In both cases, star formation activity increases towards larger clustercentric radii. The Spearman coefficient with $R/R_{200}$ is nearly identical between SAMI and TNG-Cluster ($\rho \sim 0.2$). However, the correlation with $|V_{\rm los}|/\sigma_{\rm los}$ is weaker in the simulation: while SAMI shows $\rho \sim 0.2$, the value in TNG-Cluster is $\rho \sim 0.1$.

Taken together, these results indicate that the TNG-Cluster sample successfully reproduces the main environmental trends observed in the SAMI clusters at $z=0$. Despite known differences in the absolute normalization of some galaxy properties, the dependence of $\lambda_R$, colour, and sSFR on clustercentric distance and satellite dynamics is qualitatively consistent between observations and the simulation.

This agreement provides confidence that the simulation captures the dominant environmental processes shaping satellite galaxies in clusters. In particular, it allows us to extend the analysis beyond the simulated local universe and investigate how these phase-space regions and their associated galaxy populations evolved over cosmic time. By tracing the same properties at earlier epochs, the simulation offers access to the evolutionary history of the different infall regions of clusters, providing information that is not directly accessible through observations alone.

\subsection{Evolution of medians in infalling regions}
\label{evolution infall regions}

The projected phase-space regions used in this analysis are illustrated in Fig.~\ref{fig:fig4}, where each row corresponds to a different epoch, at redshifts of $z = 0.0$, $0.2$, $0.5$, and $1.0$. The temporal evolution of galaxy properties across these regions is presented in Fig.~\ref{fig:fig5}.

The spin parameter $\lambda_R$ in Fig.~\ref{fig:fig5} displays a uniform behaviour among the infall regions A---D. In all four regions the median spin values are very similar and evolve smoothly with cosmic time, showing slightly larger values at earlier epochs. In contrast, galaxies in region E exhibit systematically lower spin values, roughly a factor of two smaller than those observed in the other regions. Despite this offset, the temporal evolution follows a similar slope.  The 16th--84th percentile bands reveal considerable overlap between regions A--D at all epochs, consistent with their similar median trends. Region E, despite its lower median, falls largely within the scatter of the other regions, indicating that the offset reflects a systematic shift in the typical spin rather than a separation into two distinct populations. Although Fig.~\ref{fig:fig5} summarizes the median trends, the full distributions of galaxies in each epoch can be seen in the projected phase-space diagrams shown in Fig.~\ref{fig:fig4}, where several systems lie above and below the median values. The spin parameter distributions for each region and epoch are also presented in the histograms of Appendix~\ref{appendix B}. This behaviour is consistent with the interpretation that galaxies in region E represent a population that has resided within the cluster potential for the longest time, accumulating the effects of dynamical processing. Processes such as tidal interactions, repeated pericentric passages, and gravitational harassment may gradually reduce the rotational support of these galaxies, leading to the lower spin values observed in this region \citep{Moore1996, Brough2017, Joshi2020, vanDeSande2021}.

The $(g-i)$ colour exhibits a clear and monotonic evolution with both time and phase-space region (Fig.~\ref{fig:fig5}). Galaxies become progressively redder toward the present day, reflecting the global decline in star formation activity with cosmic time. In addition, a well-defined gradient is present across the phase-space regions. Galaxies in region A are consistently the bluest population at all epochs, while those in region E are systematically the reddest. Intermediate regions follow a smooth transition between these two extremes. This behaviour strongly supports the interpretation that projected phase-space position traces the time since infall, with galaxies becoming progressively redder as their star formation activity declines after entering the cluster environment.

The evolution of the specific star formation rate reinforces this interpretation (Fig.~\ref{fig:fig5}). At the earliest epoch ($8~\mathrm{Gyr}$), galaxies in region A still show relatively high median sSFR values, while progressively lower values are observed toward regions B, C, and D. In contrast, galaxies located in region E are already largely quenched at this time. By $z = 0.5$, the median sSFR in all regions approaches $\log(\mathrm{sSFR} / \mathrm{yr}^{-1}) \sim -12$, indicating that a large fraction of the satellite population has already ceased star formation. This rapid decline is broadly consistent with studies linking cluster infall to delayed-then-rapid quenching and gas stripping on Gyr timescales \citep{Wetzel2013, Oman2021, Lokas2020, deIsidio2026} and with the rapid quenching sequence revealed by the individual orbital histories presented in Section~\ref{evolution individual satellites}. 

Taken together, these trends indicate that the projected phase-space structure captures the progressive transformation of infalling galaxies. While the angular momentum of satellites appears to remain largely preserved during the early stages of infall, star formation activity declines steadily, producing systematic gradients in colour and sSFR across the phase-space regions. The virialised population in region E, characterised by lower spin values and redder colours, likely represents galaxies that have spent the longest time within the cluster environment and have undergone the strongest dynamical and environmental processing. This interpretation is supported by the infall time calibration of \citet{Masson2026}, who showed that galaxies in the innermost phase-space regions typically have infall times exceeding 5--8 Gyr, consistent with the extended dynamical processing required to produce the low $\lambda_R$ values observed in region E.

\subsection{Evolution of individual infalling satellites}
\label{evolution individual satellites}

\begin{figure*}
\centering
\includegraphics{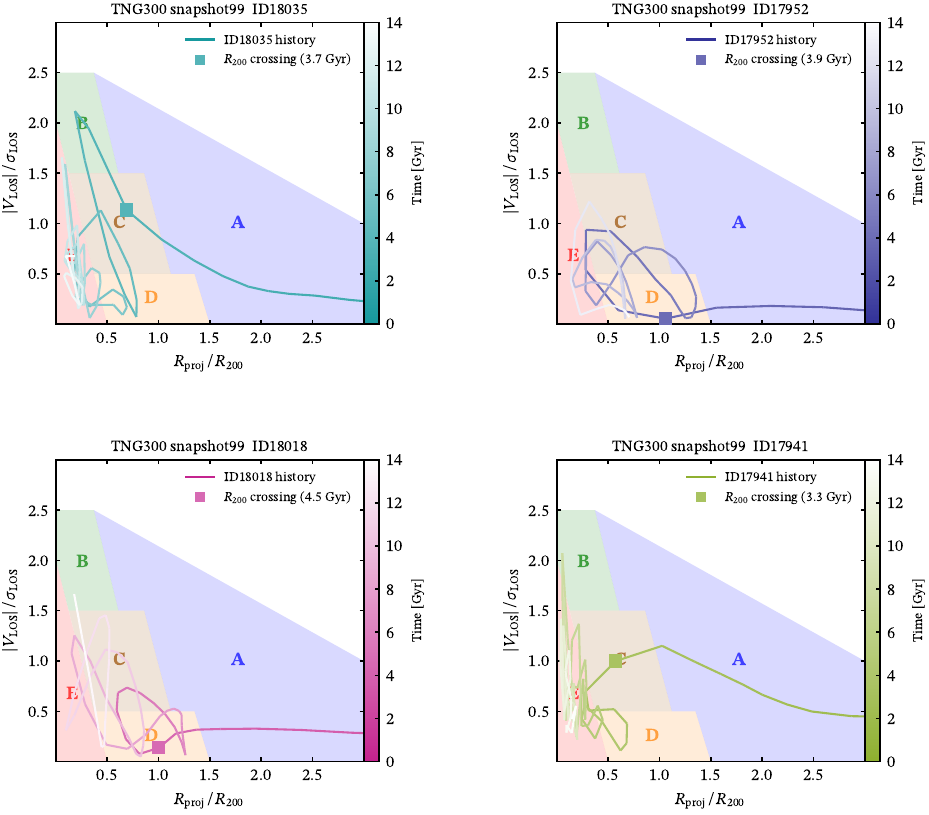}
\caption{Projected phase-space trajectories of four representative satellite galaxies selected from TNG300 at $z = 0$. Each panel shows the orbital history of one galaxy (identified by its subhalo ID) overlaid on the projected phase-space diagram, with background regions A--E following the infall-class boundaries defined by \citet{Rhee2017}. The coloured line traces the galaxy's projected phase-space position as a function of cosmic time, and the colourbar provides a qualitative indication of the early, intermediate, and late stages of each orbital trajectory. The filled square marks the epoch of first $R_{200}$ crossing, where the time indicated in parentheses in the legend refers to the cosmic time of this event; this crossing is identified as the moment at which the three-dimensional distance between the satellite and the cluster centre first falls below $R_{200}$, as can be directly seen in panel~(a) of Fig.~\ref{fig:fig7}.}
\label{fig:fig6}
\end{figure*}

\begin{figure*}
\centering
\includegraphics{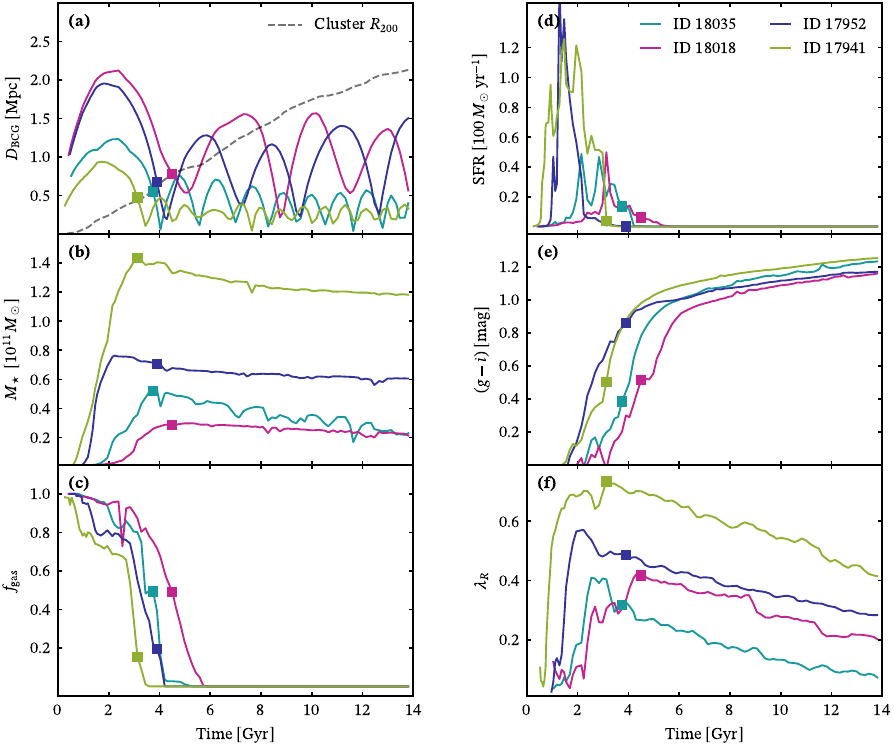}
\caption{Time evolution of physical properties for the four satellite galaxies shown in Fig.~\ref{fig:fig6}, identified by the same colour scheme. Panels show: \textbf{(a)} projected distance to the BCG ($D_\mathrm{BCG}$), with the dashed line indicating the cluster $R_{200}$; \textbf{(b)} stellar mass ($M_\star$); \textbf{(c)} gas fraction ($f_\mathrm{gas}$); \textbf{(d)} star formation rate (SFR); \textbf{(e)} optical colour $(g - i)$; and \textbf{(f)} stellar spin parameter $\lambda_R$. Filled squares mark the epoch of $R_{200}$ crossing for each galaxy, using the same colour coding as in Fig.~\ref{fig:fig6}.}
\label{fig:fig7}
\end{figure*}

The statistical trends described in the section \ref{evolution infall regions} capture the population-level behaviour of satellite galaxies across phase-space regions, but do not reveal the physical sequence of events experienced by individual infalling systems. To bridge this gap, we present in Figs.~\ref{fig:fig6} and~\ref{fig:fig7} the projected phase-space trajectories and physical evolutionary histories of four representative satellite galaxies selected from a TNG300 cluster as described in section~\ref{individual satellite}.

Figure~\ref{fig:fig6} illustrates two distinct orbital behaviours among the four satellite galaxies; the colourbar encodes cosmic time, providing a qualitative sense of the early, intermediate, and late stages of each orbital trajectory. The connection between orbital behaviour and physical evolution can be followed simultaneously in Fig.~\ref{fig:fig7}: the pericentric passages are directly visible as repeated oscillations in the distance to the BCG shown in panel~(a), while in Fig.~\ref{fig:fig6} the same passages correspond to excursions into the innermost phase-space regions. Galaxies ID18035 and ID17941 enter the cluster and rapidly sink into the inner cluster regions, spending the majority of their subsequent evolution in region E and undergoing numerous close pericentric passages near the BCG, individually traceable as the repeated oscillations in panel~(a) of Fig.~\ref{fig:fig7}. In contrast, ID17952 and ID18018 follow more extended orbits with larger apocentric distances, completing approximately four pericentric passages and spending significantly less time in the dense cluster core, traversing the intermediate phase-space regions more persistently even at late times.

The sequence of gas stripping and quenching is broadly consistent across all four galaxies. In all cases, the first crossing of $R_{200}$ coincides with the onset of rapid gas stripping: $f_\mathrm{gas}$ drops sharply within a few Gyr of infall, accompanied by a simultaneous decline in SFR and subsequent reddening of $(g-i)$. This sequence is consistent with ram pressure stripping efficiently removing the cold gas reservoir upon infall, triggering rapid environmental quenching. The spin parameter $\lambda_R$, however, responds on markedly longer timescales. Following $R_{200}$ crossing, $\lambda_R$ enters a gradual, approximately linear decline that persists to $z = 0$, with the magnitude and pace of this decline depending on the orbital history of each satellite.

Comparing ID18035 and ID18018, which converge to a similar stellar mass of $\sim 0.2 \times 10^{11}\,\mathrm{M_\odot}$ at $z = 0$ and reach comparable peak $\lambda_R$ values of $\sim 0.4$ at the onset of infall, illustrates this clearly: despite these similarities, ID18035, with its deeply embedded orbit and extended residence in region E, experiences a $\lambda_R$ decline approximately 30 per cent greater than that of ID18018, whose more extended orbit limits its exposure to the dense cluster core. This comparison highlights that the time spent in region E, and thus the cumulative exposure to tidal interactions, gravitational harassment by the denser satellite population, and repeated pericentric passages \citep{Moore1996, Joshi2020, Lokas2020}, appears to be associated with $\lambda_R$ decline. Stellar mass acts simultaneously as an important modulating factor. Among the two deeply embedded galaxies, ID17941 ($\sim 1.4 \times 10^{11}\,\mathrm{M_\odot}$) and ID18035 ($\sim 0.4 \times 10^{11}\,\mathrm{M_\odot}$), which spend comparable amounts of time in region E, the higher-mass system retains significantly more of its rotational support, declining by $\sim 40$ per cent compared to $\sim 75$ per cent for the lower-mass counterpart. This suggests that more massive stellar discs are intrinsically better able to resist the dynamical heating and perturbations that pervade the virialised cluster core.

These individual histories provide a physically coherent picture that directly complements the statistical results of Section~\ref{evolution infall regions}. The statistical evolution shown in Fig.~\ref{fig:fig5} covers the last $\sim 8$ Gyr, a period during which most satellites are already in an advanced stage of environmental processing. The individual histories in Fig.~\ref{fig:fig7} reveal what preceded this period: the pre-infall star formation peak, the abrupt gas stripping at $R_{200}$ crossing, and the initiation of the gradual spin parameter decline. Together, these two perspectives demonstrate that while quenching of star formation is rapid and closely tied to the moment of infall, the dynamical transformation of the stellar component, as traced by $\lambda_R$, is a slower and more gradual process driven by the cumulative exposure to the dynamical processes that pervade the virialised cluster core.  This timescale mismatch offers an explanation for the weak $\lambda_R$ gradient observed across regions A--D in Fig.~\ref{fig:fig5}: since the spin decline operates over several Gyr, satellites traversing these regions on their first infall do not reside there long enough for the effect to become statistically detectable, whereas the extended residence of ancient infallers in region E provides the cumulative exposure necessary for the signal to emerge. 

The gradual nature of this decline is also consistent with starvation acting as a long-term background mechanism: once the hot gas reservoir is stripped upon infall, the galaxy continues to consume its remaining cold gas over several Gyr without experiencing significant kinematic perturbation, leaving the stellar disc largely undisturbed while star formation slowly fades \citep{Larson1980, Trussler2020, deIsidio2026}. In this picture, the slow monotonic decline of $\lambda_R$ reflects not only the cumulative dynamical effects of the cluster potential, but also the absence of the disruptive kinematic imprints on $\lambda_R$ that would otherwise accompany more violent quenching channels.

\subsection{Environmental processing in cluster cores}
\label{cluster cores}

The preferential transformation of satellites in the virialised core suggests that the cluster potential itself, rather than infall alone, is the dominant driver of dynamical evolution. This raises the question of which specific processes are responsible, and on what timescales they operate.

The timescales associated with these transformations are also relevant for interpreting our results. \citet{Masson2026} used TNG300 and TNG-Cluster to derive a calibration of galaxy infall times from projected phase-space position and stellar mass, providing a framework to connect the phase-space regions defined by \citet{Rhee2017} to quantitative infall time estimates. Their results confirm that galaxies in the innermost phase-space regions, corresponding to our region E, have the longest infall times, typically exceeding $5$--$8$ Gyr. This is entirely consistent with the gradual decline of $\lambda_R$ we observe in our individual orbital histories, where the full suppression of rotational support requires several Gyr of repeated dynamical interactions within the cluster core. Together, these results reinforce the interpretation that the observed $\lambda_R$ gradient across phase-space regions is not simply a snapshot effect, but reflects the cumulative dynamical history of satellite galaxies since their first infall.

This picture is broadly consistent with the findings of \citet{Lokas2020}, who studied the tidal evolution of galaxies in the most massive cluster of TNG100 and identified three distinct evolutionary classes based on the number of pericentric passages: infalling, weakly evolved, and strongly evolved systems. Their strongly evolved sample, which experienced multiple pericentric passages and the highest integrated tidal forces, showed the most significant suppression of ordered rotation, with the rotation parameter $f$ dropping to near zero in the most extreme cases, alongside morphological transformation from oblate rotating disks to prolate, bar-like spheroids. This is qualitatively consistent with the low $\lambda_R$ values we find for galaxies in region E of our phase-space diagrams, which are predominantly systems that have undergone multiple pericentric passages within the cluster core.

An additional layer of complexity arises from the possibility that part of this dynamical transformation may have occurred prior to cluster infall. The role of pre-processing in group environments prior to cluster infall adds a further layer of complexity to the interpretation of phase-space position as a proxy for infall history. The dynamical interactions responsible for the suppression of stellar angular momentum in cluster cores also operate in lower-mass group environments, albeit at reduced velocities and over longer timescales \citep{Lokas2023}. Since galaxies may spend several Gyr in group-scale halos before being accreted by a cluster, a fraction of the $\lambda_R$ decline observed in the virialised population could in principle have been initiated prior to cluster infall. Investigating the environmental dependence of $\lambda_R$ in group-scale halos, where these processes operate more gently and over extended periods, represents a natural extension of this work and would help disentangle the relative contributions of pre-processing and cluster-driven dynamical transformation to the observed angular momentum gradients across projected phase-space.

\section{Conclusions}
\label{conclusions}
In this work we investigated the environmental dependence and cosmic evolution of satellite galaxy properties in clusters by combining observations from the SAMI Galaxy Survey with predictions from the TNG-Cluster cosmological simulation. Using projected phase-space diagrams, we analysed how the stellar spin parameter ($\lambda_R$), $(g-i)$ colour, and sSFR vary as a function of clustercentric distance and line-of-sight velocity. By tracing the same cluster systems across cosmic time and following the orbital and physical histories of individual satellite galaxies, we connected present-day observed trends to the physical processes driving galaxy transformation during infall. Our main results can be summarised as follows:

\begin{enumerate}
\item \textbf{Consistency between observations and simulations.} 
The environmental trends observed in the SAMI clusters are qualitatively reproduced by the TNG-Cluster simulation at $z = 0$. In both datasets, $\lambda_R$, $(g-i)$ colour, and sSFR show statistically consistent correlations with projected clustercentric distance and satellite velocity, validating the use of TNG-Cluster as a framework to interpret and extend the observational results. For \(\lambda_R\), this agreement concerns the environmental trends rather than the absolute normalisation of the spin measurements.

\item \textbf{Projected phase-space as a tracer of infall history.} 
The distribution of galaxy properties across projected phase-space regions, together with their temporal evolution with redshift, supports the interpretation that these regions trace different stages of the infall process. Galaxies in outer regions associated with recent infall exhibit higher star formation activity and bluer colours, while galaxies in the virialised region are typically redder and more quenched, indicating longer residence times within the cluster potential.

\item \textbf{Environmental quenching after infall.}
Star formation quenching is rapid and closely tied to the moment of $R_{200}$ crossing. Individual orbital histories reveal that $f_\mathrm{gas}$ drops sharply within a few Gyr of infall, accompanied by a simultaneous decline in SFR and subsequent reddening of $(g-i)$, consistent with ram pressure stripping efficiently removing the cold gas reservoir upon entry into the cluster potential.  Over longer timescales, starvation further sustains this quenching by preventing the replenishment of the cold gas reservoir, consolidating the transition to a passive state.

\item \textbf{Slow and cumulative angular momentum evolution.} 
In contrast to the rapid quenching of star formation, the suppression of $\lambda_R$ operates on markedly longer timescales, entering a gradual decline following $R_{200}$ crossing that persists to $z = 0$. Galaxies in region E exhibit systematically lower spin values than outer infall regions,reflecting cumulative dynamical processing within the cluster core. The absence of a comparable gradient across regions A--D is consistent with the slow pace of this transformation, which exceeds the typical residence time of satellites in any single phase-space region prior to  virialisation. The gradual and undisturbed nature of this decline further suggests that starvation acts as a long-term background mechanism operating alongside the dynamical effects of the cluster potential.

\item \textbf{Orbital history and stellar mass modulate $\lambda_R$ decline.}
Individual orbital histories reveal that the magnitude of $\lambda_R$ suppression is governed by the interplay between two factors. The time spent in the virialised cluster core is the primary driver, as galaxies with deeply embedded orbits experience a significantly greater decline in rotational support than those with more extended orbits. Stellar mass acts as a simultaneous modulating factor, with more massive discs better able to resist the dynamical heating associated with the cluster environment.

\end{enumerate}

The combination of statistical phase-space analysis and individual orbital histories provides a comprehensive across-redshift picture of this transformation. While the correlation between $\lambda_R$ and environmental indicators remains weak in observational samples such as SAMI \citep{vandeSande2017a, Santucci2023} --- a signal reproduced with similar strength in TNG-Cluster at $z = 0$ --- the tracking of individual satellites reveals a more direct connection between the cluster core environment and the evolution of stellar angular momentum. The gradual and sustained decline of $\lambda_R$ following infall, modulated by orbital history and stellar mass, suggests that the weak statistical signal reflects the cumulative nature of dynamical transformation rather than an absence of environmental influence: the spin decline operates on timescales that exceed the typical residence time of satellites in any single outer infall region, making the effect statistically detectable only in the virialised core where ancient infallers have accumulated sufficient dynamical exposure. This stands in marked contrast with the rapid quenching of the gaseous component, which leaves an imprint across all infall regions.

\section*{Acknowledgements}

GG acknowledges the European Southern Observatory (ESO) for the opportunity to participate in the 2025 Summer Research Programme, during which this work was carried out.

GG acknowledges support from \textit{Coordenac\c c\~ao de Aperfei\c coamento de Pessoal de N\'ivel Superior} -- Brasil (CAPES) -- Finance Code 001, and from the Brazilian agency \textit{Conselho Nacional de Desenvolvimento Cient\'ifico e Tecnol\'ogico} (CNPq). GG thanks the IllustrisTNG collaboration for generously providing access to IllustrisTNG simulation data and computational resources via the online JupyterLab workspace. RM acknowledges support from CNPq through grants 406908/2018-4 and 307205/2021-5, and from \textit{Funda\c c\~ao de Apoio \`a Ci\^encia, Tecnologia e Inova\c c\~ao do Paran\'a} through grant 18.148.096-3 -- NAPI \textit{Fen\^omenos Extremos do Universo}. RPA acknowledges financial support from FAPESP under grant 2024/13224-9. IM acknowledges support from the European Research Council (ERC) under the European Union's Horizon Europe research and innovation programme ERC CoG (Grant agreement No. 101045437). NdI gratefully acknowledges the IMPRS program and ESO for the support and funding of his PhD.

\section*{Data availability}
The data underlying this article will be shared on reasonable request to the corresponding author.

\bibliographystyle{mnras.bst}
\bibliography{paper.bib}

@ARTICLE{Dressler84,
       author = {{Dressler}, A.},
        title = "{The Evolution of Galaxies in Clusters}",
      journal = {\araa},
         year = 1984,
        month = jan,
       volume = {22},
        pages = {185-222},
          doi = {10.1146/annurev.astro.22.1.185},
       adsurl = {https://ui.adsabs.harvard.edu/abs/1984ARA&A..22..185D}
}

@ARTICLE{LaraLopez22,
       author = {{Lara-L{\'o}pez}, M.~A. and {Gal{\'a}n-de Anta}, P.~M. and {Sarzi}, M. and {Iodice}, E. and {Davis}, T.~A. and {Zabel}, N. and {Corsini}, E.~M. and {de Zeeuw}, P.~T. and {Fahrion}, K. and {Falc{\'o}n-Barroso}, J. and {Gadotti}, D.~A. and {McDermid}, R.~M. and {Pinna}, F. and {Rodriguez-Gomez}, V. and {van de Ven}, G. and {Zhu}, L. and {Coccato}, L. and {Lyubenova}, M. and {Mart{\'\i}n-Navarro}, I.},
        title = "{The Fornax3D project: The environmental impact on gas metallicity gradients in Fornax cluster galaxies}",
      journal = {\aap},
         year = 2022,
        month = apr,
       volume = {660},
          eid = {A105},
        pages = {A105},
          doi = {10.1051/0004-6361/202142790},
archivePrefix = {arXiv},
       eprint = {2202.04128},
 primaryClass = {astro-ph.GA},
       adsurl = {https://ui.adsabs.harvard.edu/abs/2022A&A...660A.105L}
}

@ARTICLE{Goubert25,
       author = {{Goubert}, Paul H. and {Bluck}, Asa F.~L. and {Piotrowska}, Joanna M. and {Torrey}, Paul and {Maiolino}, Roberto and {Franco}, Thomas Pinto and {Casimiro}, Camilo and {Cea}, Nicolas},
        title = "{Environmental versus intrinsic quenching at cosmic noon: predictions from cosmological hydrodynamical simulations for VLT-MOONRISE}",
      journal = {\mnras},
         year = 2025,
        month = nov,
       volume = {543},
       number = {3},
        pages = {2006-2034},
          doi = {10.1093/mnras/staf1554},
archivePrefix = {arXiv},
       eprint = {2509.09626},
 primaryClass = {astro-ph.GA},
       adsurl = {https://ui.adsabs.harvard.edu/abs/2025MNRAS.543.2006G}
}

@ARTICLE{Owers2017,
       author = {{Owers}, M.~S. and {Allen}, J.~T. and {Baldry}, I. and {Bryant}, J.~J. and {Cecil}, G.~N. and {Cortese}, L. and {Croom}, S.~M. and {Driver}, S.~P. and {Fogarty}, L.~M.~R. and {Green}, A.~W. and {Helmich}, E. and {de Jong}, J.~T.~A. and {Kuijken}, K. and {Mahajan}, S. and {McFarland}, J. and {Pracy}, M.~B. and {Robotham}, A.~G.~S. and {Sikkema}, G. and {Sweet}, S. and {Taylor}, E.~N. and {Verdoes Kleijn}, G. and {Bauer}, A.~E. and {Bland-Hawthorn}, J. and {Brough}, S. and {Colless}, M. and {Couch}, W.~J. and {Davies}, R.~L. and {Drinkwater}, M.~J. and {Goodwin}, M. and {Hopkins}, A.~M. and {Konstantopoulos}, I.~S. and {Foster}, C. and {Lawrence}, J.~S. and {Lorente}, N.~P.~F. and {Medling}, A.~M. and {Metcalfe}, N. and {Richards}, S.~N. and {van de Sande}, J. and {Scott}, N. and {Shanks}, T. and {Sharp}, R. and {Thomas}, A.~D. and {Tonini}, C.},
        title = "{The SAMI Galaxy Survey: the cluster redshift survey, target selection and cluster properties}",
      journal = {\mnras},
         year = 2017,
        month = jun,
       volume = {468},
       number = {2},
        pages = {1824-1849},
          doi = {10.1093/mnras/stx562},
archivePrefix = {arXiv},
       eprint = {1703.00997},
 primaryClass = {astro-ph.GA},
       adsurl = {https://ui.adsabs.harvard.edu/abs/2017MNRAS.468.1824O}
}

@ARTICLE{Pallero2025,
       author = {{Pallero}, Diego and {Galaz}, Gaspar and {Tissera}, Patricia B. and {G{\'o}mez}, Facundo A. and {Monachesi}, Antonela and {Sif{\'o}n}, Cristobal and {Tapia-Contreras}, Brian},
        title = "{The formation and evolution of supermassive disks in IllustrisTNG}",
      journal = {\aap},
         year = 2025,
        month = jul,
       volume = {699},
          eid = {A376},
        pages = {A376},
          doi = {10.1051/0004-6361/202554509},
archivePrefix = {arXiv},
       eprint = {2507.00141},
 primaryClass = {astro-ph.GA},
       adsurl = {https://ui.adsabs.harvard.edu/abs/2025A&A...699A.376P}
}

@ARTICLE{Lagos2017,
       author = {{Lagos}, Claudia del P. and {Theuns}, Tom and {Stevens}, Adam R.~H. and {Cortese}, Luca and {Padilla}, Nelson D. and {Davis}, Timothy A. and {Contreras}, Sergio and {Croton}, Darren},
        title = "{Angular momentum evolution of galaxies in EAGLE}",
      journal = {\mnras},
         year = 2017,
        month = feb,
       volume = {464},
       number = {4},
        pages = {3850-3870},
          doi = {10.1093/mnras/stw2610},
archivePrefix = {arXiv},
       eprint = {1609.01739},
 primaryClass = {astro-ph.GA},
       adsurl = {https://ui.adsabs.harvard.edu/abs/2017MNRAS.464.3850L}
}

@ARTICLE{Springel2010,
       author = {{Springel}, Volker},
        title = "{E pur si muove: Galilean-invariant cosmological hydrodynamical simulations on a moving mesh}",
      journal = {\mnras},
         year = 2010,
        month = jan,
       volume = {401},
       number = {2},
        pages = {791-851},
          doi = {10.1111/j.1365-2966.2009.15715.x},
archivePrefix = {arXiv},
       eprint = {0901.4107},
 primaryClass = {astro-ph.CO},
       adsurl = {https://ui.adsabs.harvard.edu/abs/2010MNRAS.401..791S}
}

@ARTICLE{Weinberger2017,
       author = {{Weinberger}, Rainer and {Springel}, Volker and {Hernquist}, Lars and {Pillepich}, Annalisa and {Marinacci}, Federico and {Pakmor}, R{\"u}diger and {Nelson}, Dylan and {Genel}, Shy and {Vogelsberger}, Mark and {Naiman}, Jill and {Torrey}, Paul},
        title = "{Simulating galaxy formation with black hole driven thermal and kinetic feedback}",
      journal = {\mnras},
         year = 2017,
        month = mar,
       volume = {465},
       number = {3},
        pages = {3291-3308},
          doi = {10.1093/mnras/stw2944},
archivePrefix = {arXiv},
       eprint = {1607.03486},
 primaryClass = {astro-ph.GA},
       adsurl = {https://ui.adsabs.harvard.edu/abs/2017MNRAS.465.3291W}
}

@ARTICLE{Pillepich2018a,
       author = {{Pillepich}, Annalisa and {Springel}, Volker and {Nelson}, Dylan and {Genel}, Shy and {Naiman}, Jill and {Pakmor}, R{\"u}diger and {Hernquist}, Lars and {Torrey}, Paul and {Vogelsberger}, Mark and {Weinberger}, Rainer and {Marinacci}, Federico},
        title = "{Simulating galaxy formation with the IllustrisTNG model}",
      journal = {\mnras},
         year = 2018,
        month = jan,
       volume = {473},
       number = {3},
        pages = {4077-4106},
          doi = {10.1093/mnras/stx2656},
archivePrefix = {arXiv},
       eprint = {1703.02970},
 primaryClass = {astro-ph.GA},
       adsurl = {https://ui.adsabs.harvard.edu/abs/2018MNRAS.473.4077P}
}

@ARTICLE{Springel2018,
       author = {{Springel}, Volker and {Pakmor}, R{\"u}diger and {Pillepich}, Annalisa and {Weinberger}, Rainer and {Nelson}, Dylan and {Hernquist}, Lars and {Vogelsberger}, Mark and {Genel}, Shy and {Torrey}, Paul and {Marinacci}, Federico and {Naiman}, Jill},
        title = "{First results from the IllustrisTNG simulations: matter and galaxy clustering}",
      journal = {\mnras},
         year = 2018,
        month = mar,
       volume = {475},
       number = {1},
        pages = {676-698},
          doi = {10.1093/mnras/stx3304},
archivePrefix = {arXiv},
       eprint = {1707.03397},
 primaryClass = {astro-ph.GA},
       adsurl = {https://ui.adsabs.harvard.edu/abs/2018MNRAS.475..676S}
}

@ARTICLE{Naiman2018,
       author = {{Naiman}, Jill P. and {Pillepich}, Annalisa and {Springel}, Volker and {Ramirez-Ruiz}, Enrico and {Torrey}, Paul and {Vogelsberger}, Mark and {Pakmor}, R{\"u}diger and {Nelson}, Dylan and {Marinacci}, Federico and {Hernquist}, Lars and {Weinberger}, Rainer and {Genel}, Shy},
        title = "{First results from the IllustrisTNG simulations: a tale of two elements - chemical evolution of magnesium and europium}",
      journal = {\mnras},
         year = 2018,
        month = jun,
       volume = {477},
       number = {1},
        pages = {1206-1224},
          doi = {10.1093/mnras/sty618},
archivePrefix = {arXiv},
       eprint = {1707.03401},
 primaryClass = {astro-ph.GA},
       adsurl = {https://ui.adsabs.harvard.edu/abs/2018MNRAS.477.1206N}
}

@ARTICLE{Marinacci2018,
       author = {{Marinacci}, Federico and {Vogelsberger}, Mark and {Pakmor}, R{\"u}diger and {Torrey}, Paul and {Springel}, Volker and {Hernquist}, Lars and {Nelson}, Dylan and {Weinberger}, Rainer and {Pillepich}, Annalisa and {Naiman}, Jill and {Genel}, Shy},
        title = "{First results from the IllustrisTNG simulations: radio haloes and magnetic fields}",
      journal = {\mnras},
         year = 2018,
        month = nov,
       volume = {480},
       number = {4},
        pages = {5113-5139},
          doi = {10.1093/mnras/sty2206},
archivePrefix = {arXiv},
       eprint = {1707.03396},
 primaryClass = {astro-ph.CO},
       adsurl = {https://ui.adsabs.harvard.edu/abs/2018MNRAS.480.5113M}
}

@ARTICLE{Nelson2018,
       author = {{Nelson}, Dylan and {Pillepich}, Annalisa and {Springel}, Volker and {Weinberger}, Rainer and {Hernquist}, Lars and {Pakmor}, R{\"u}diger and {Genel}, Shy and {Torrey}, Paul and {Vogelsberger}, Mark and {Kauffmann}, Guinevere and {Marinacci}, Federico and {Naiman}, Jill},
        title = "{First results from the IllustrisTNG simulations: the galaxy colour bimodality}",
      journal = {\mnras},
         year = 2018,
        month = mar,
       volume = {475},
       number = {1},
        pages = {624-647},
          doi = {10.1093/mnras/stx3040},
archivePrefix = {arXiv},
       eprint = {1707.03395},
 primaryClass = {astro-ph.GA},
       adsurl = {https://ui.adsabs.harvard.edu/abs/2018MNRAS.475..624N}
}

@ARTICLE{Nelson2024,
       author = {{Nelson}, Dylan and {Pillepich}, Annalisa and {Ayromlou}, Mohammadreza and {Lee}, Wonki and {Lehle}, Katrin and {Rohr}, Eric and {Truong}, Nhut},
        title = "{Introducing the TNG-Cluster simulation: Overview and the physical properties of the gaseous intracluster medium}",
      journal = {\aap},
         year = 2024,
        month = jun,
       volume = {686},
          eid = {A157},
        pages = {A157},
          doi = {10.1051/0004-6361/202348608},
archivePrefix = {arXiv},
       eprint = {2311.06338},
 primaryClass = {astro-ph.GA},
       adsurl = {https://ui.adsabs.harvard.edu/abs/2024A&A...686A.157N}
}

@ARTICLE{Planck2016,
       author = {{Planck Collaboration} and {Ade}, P.~A.~R. and {Aghanim}, N. and {Arnaud}, M. and {Ashdown}, M. and {Aumont}, J. and {Baccigalupi}, C. and {Banday}, A.~J. and {Barreiro}, R.~B. and {Bartlett}, J.~G. and {Bartolo}, N. and {Battaner}, E. and {Battye}, R. and {Benabed}, K. and {Beno{\^\i}t}, A. and {Benoit-L{\'e}vy}, A. and {Bernard}, J.-P. and {Bersanelli}, M. and {Bielewicz}, P. and {Bock}, J.~J. and {Bonaldi}, A. and {Bonavera}, L. and {Bond}, J.~R. and {Borrill}, J. and {Bouchet}, F.~R. and {Boulanger}, F. and {Bucher}, M. and {Burigana}, C. and {Butler}, R.~C. and {Calabrese}, E. and {Cardoso}, J.-F. and {Catalano}, A. and {Challinor}, A. and {Chamballu}, A. and {Chary}, R.-R. and {Chiang}, H.~C. and {Chluba}, J. and {Christensen}, P.~R. and {Church}, S. and {Clements}, D.~L. and {Colombi}, S. and {Colombo}, L.~P.~L. and {Combet}, C. and {Coulais}, A. and {Crill}, B.~P. and {Curto}, A. and {Cuttaia}, F. and {Danese}, L. and {Davies}, R.~D. and {Davis}, R.~J. and {de Bernardis}, P. and {de Rosa}, A. and {de Zotti}, G. and {Delabrouille}, J. and {D{\'e}sert}, F.-X. and {Di Valentino}, E. and {Dickinson}, C. and {Diego}, J.~M. and {Dolag}, K. and {Dole}, H. and {Donzelli}, S. and {Dor{\'e}}, O. and {Douspis}, M. and {Ducout}, A. and {Dunkley}, J. and {Dupac}, X. and {Efstathiou}, G. and {Elsner}, F. and {En{\ss}lin}, T.~A. and {Eriksen}, H.~K. and {Farhang}, M. and {Fergusson}, J. and {Finelli}, F. and {Forni}, O. and {Frailis}, M. and {Fraisse}, A.~A. and {Franceschi}, E. and {Frejsel}, A. and {Galeotta}, S. and {Galli}, S. and {Ganga}, K. and {Gauthier}, C. and {Gerbino}, M. and {Ghosh}, T. and {Giard}, M. and {Giraud-H{\'e}raud}, Y. and {Giusarma}, E. and {Gjerl{\o}w}, E. and {Gonz{\'a}lez-Nuevo}, J. and {G{\'o}rski}, K.~M. and {Gratton}, S. and {Gregorio}, A. and {Gruppuso}, A. and {Gudmundsson}, J.~E. and {Hamann}, J. and {Hansen}, F.~K. and {Hanson}, D. and {Harrison}, D.~L. and {Helou}, G. and {Henrot-Versill{\'e}}, S. and {Hern{\'a}ndez-Monteagudo}, C. and {Herranz}, D. and {Hildebrandt}, S.~R. and {Hivon}, E. and {Hobson}, M. and {Holmes}, W.~A. and {Hornstrup}, A. and {Hovest}, W. and {Huang}, Z. and {Huffenberger}, K.~M. and {Hurier}, G. and {Jaffe}, A.~H. and {Jaffe}, T.~R. and {Jones}, W.~C. and {Juvela}, M. and {Keih{\"a}nen}, E. and {Keskitalo}, R. and {Kisner}, T.~S. and {Kneissl}, R. and {Knoche}, J. and {Knox}, L. and {Kunz}, M. and {Kurki-Suonio}, H. and {Lagache}, G. and {L{\"a}hteenm{\"a}ki}, A. and {Lamarre}, J.-M. and {Lasenby}, A. and {Lattanzi}, M. and {Lawrence}, C.~R. and {Leahy}, J.~P. and {Leonardi}, R. and {Lesgourgues}, J. and {Levrier}, F. and {Lewis}, A. and {Liguori}, M. and {Lilje}, P.~B. and {Linden-V{\o}rnle}, M. and {L{\'o}pez-Caniego}, M. and {Lubin}, P.~M. and {Mac{\'\i}as-P{\'e}rez}, J.~F. and {Maggio}, G. and {Maino}, D. and {Mandolesi}, N. and {Mangilli}, A. and {Marchini}, A. and {Maris}, M. and {Martin}, P.~G. and {Martinelli}, M. and {Mart{\'\i}nez-Gonz{\'a}lez}, E. and {Masi}, S. and {Matarrese}, S. and {McGehee}, P. and {Meinhold}, P.~R. and {Melchiorri}, A. and {Melin}, J.-B. and {Mendes}, L. and {Mennella}, A. and {Migliaccio}, M. and {Millea}, M. and {Mitra}, S. and {Miville-Desch{\^e}nes}, M.-A. and {Moneti}, A. and {Montier}, L. and {Morgante}, G. and {Mortlock}, D. and {Moss}, A. and {Munshi}, D. and {Murphy}, J.~A. and {Naselsky}, P. and {Nati}, F. and {Natoli}, P. and {Netterfield}, C.~B. and {N{\o}rgaard-Nielsen}, H.~U. and {Noviello}, F. and {Novikov}, D. and {Novikov}, I. and {Oxborrow}, C.~A. and {Paci}, F. and {Pagano}, L. and {Pajot}, F. and {Paladini}, R. and {Paoletti}, D. and {Partridge}, B. and {Pasian}, F. and {Patanchon}, G. and {Pearson}, T.~J. and {Perdereau}, O. and {Perotto}, L. and {Perrotta}, F. and {Pettorino}, V. and {Piacentini}, F. and {Piat}, M. and {Pierpaoli}, E. and {Pietrobon}, D. and {Plaszczynski}, S. and {Pointecouteau}, E. and {Polenta}, G. and {Popa}, L. and {Pratt}, G.~W. and {Pr{\'e}zeau}, G.},
        title = "{Planck 2015 results. XIII. Cosmological parameters}",
      journal = {\aap},
         year = 2016,
        month = sep,
       volume = {594},
          eid = {A13},
        pages = {A13},
          doi = {10.1051/0004-6361/201525830},
archivePrefix = {arXiv},
       eprint = {1502.01589},
 primaryClass = {astro-ph.CO},
       adsurl = {https://ui.adsabs.harvard.edu/abs/2016A&A...594A..13P}
}

@ARTICLE{Springel2001,
       author = {{Springel}, Volker and {White}, Simon D.~M. and {Tormen}, Giuseppe and {Kauffmann}, Guinevere},
        title = "{Populating a cluster of galaxies - I. Results at z=0}",
      journal = {\mnras},
         year = 2001,
        month = dec,
       volume = {328},
       number = {3},
        pages = {726-750},
          doi = {10.1046/j.1365-8711.2001.04912.x},
archivePrefix = {arXiv},
       eprint = {astro-ph/0012055},
 primaryClass = {astro-ph},
       adsurl = {https://ui.adsabs.harvard.edu/abs/2001MNRAS.328..726S}
}

@ARTICLE{Emsellem2007,
       author = {{Emsellem}, Eric and {Cappellari}, Michele and {Krajnovi{\'c}}, Davor and {van de Ven}, Glenn and {Bacon}, R. and {Bureau}, M. and {Davies}, Roger L. and {de Zeeuw}, P.~T. and {Falc{\'o}n-Barroso}, Jes{\'u}s and {Kuntschner}, Harald and {McDermid}, Richard and {Peletier}, Reynier F. and {Sarzi}, Marc},
        title = "{The SAURON project - IX. A kinematic classification for early-type galaxies}",
      journal = {\mnras},
         year = 2007,
        month = aug,
       volume = {379},
       number = {2},
        pages = {401-417},
          doi = {10.1111/j.1365-2966.2007.11752.x},
archivePrefix = {arXiv},
       eprint = {astro-ph/0703531},
 primaryClass = {astro-ph},
       adsurl = {https://ui.adsabs.harvard.edu/abs/2007MNRAS.379..401E}
}

@ARTICLE{vandeSande2017a,
       author = {{van de Sande}, Jesse and {Bland-Hawthorn}, Joss and {Fogarty}, Lisa M.~R. and {Cortese}, Luca and {d'Eugenio}, Francesco and {Croom}, Scott M. and {Scott}, Nicholas and {Allen}, James T. and {Brough}, Sarah and {Bryant}, Julia J. and {Cecil}, Gerald and {Colless}, Matthew and {Couch}, Warrick J. and {Davies}, Roger and {Elahi}, Pascal J. and {Foster}, Caroline and {Goldstein}, Gregory and {Goodwin}, Michael and {Groves}, Brent and {Ho}, I.-Ting and {Jeong}, Hyunjin and {Jones}, D. Heath and {Konstantopoulos}, Iraklis S. and {Lawrence}, Jon S. and {Leslie}, Sarah K. and {L{\'o}pez-S{\'a}nchez}, {\'A}ngel R. and {McDermid}, Richard M. and {McElroy}, Rebecca and {Medling}, Anne M. and {Oh}, Sree and {Owers}, Matt S. and {Richards}, Samuel N. and {Schaefer}, Adam L. and {Sharp}, Rob and {Sweet}, Sarah M. and {Taranu}, Dan and {Tonini}, Chiara and {Walcher}, C. Jakob and {Yi}, Sukyoung K.},
        title = "{The SAMI Galaxy Survey: Revisiting Galaxy Classification through High-order Stellar Kinematics}",
      journal = {\apj},
         year = 2017,
        month = jan,
       volume = {835},
       number = {1},
          eid = {104},
        pages = {104},
          doi = {10.3847/1538-4357/835/1/104},
archivePrefix = {arXiv},
       eprint = {1611.07039},
 primaryClass = {astro-ph.GA},
       adsurl = {https://ui.adsabs.harvard.edu/abs/2017ApJ...835..104V}
}

@ARTICLE{Bryant2015,
       author = {{Bryant}, J.~J. and {Owers}, M.~S. and {Robotham}, A.~S.~G. and {Croom}, S.~M. and {Driver}, S.~P. and {Drinkwater}, M.~J. and {Lorente}, N.~P.~F. and {Cortese}, L. and {Scott}, N. and {Colless}, M. and {Schaefer}, A. and {Taylor}, E.~N. and {Konstantopoulos}, I.~S. and {Allen}, J.~T. and {Baldry}, I. and {Barnes}, L. and {Bauer}, A.~E. and {Bland-Hawthorn}, J. and {Bloom}, J.~V. and {Brooks}, A.~M. and {Brough}, S. and {Cecil}, G. and {Couch}, W. and {Croton}, D. and {Davies}, R. and {Ellis}, S. and {Fogarty}, L.~M.~R. and {Foster}, C. and {Glazebrook}, K. and {Goodwin}, M. and {Green}, A. and {Gunawardhana}, M.~L. and {Hampton}, E. and {Ho}, I.-T. and {Hopkins}, A.~M. and {Kewley}, L. and {Lawrence}, J.~S. and {Leon-Saval}, S.~G. and {Leslie}, S. and {McElroy}, R. and {Lewis}, G. and {Liske}, J. and {L{\'o}pez-S{\'a}nchez}, {\'A}. R. and {Mahajan}, S. and {Medling}, A.~M. and {Metcalfe}, N. and {Meyer}, M. and {Mould}, J. and {Obreschkow}, D. and {O'Toole}, S. and {Pracy}, M. and {Richards}, S.~N. and {Shanks}, T. and {Sharp}, R. and {Sweet}, S.~M. and {Thomas}, A.~D. and {Tonini}, C. and {Walcher}, C.~J.},
        title = "{The SAMI Galaxy Survey: instrument specification and target selection}",
      journal = {\mnras},
         year = 2015,
        month = mar,
       volume = {447},
       number = {3},
        pages = {2857-2879},
          doi = {10.1093/mnras/stu2635},
archivePrefix = {arXiv},
       eprint = {1407.7335},
 primaryClass = {astro-ph.GA},
       adsurl = {https://ui.adsabs.harvard.edu/abs/2015MNRAS.447.2857B}
}

@ARTICLE{Gunn1972,
       author = {{Gunn}, James E. and {Gott}, III, J. Richard},
        title = "{On the Infall of Matter Into Clusters of Galaxies and Some Effects on Their Evolution}",
      journal = {\apj},
         year = 1972,
        month = aug,
       volume = {176},
        pages = {1},
          doi = {10.1086/151605},
       adsurl = {https://ui.adsabs.harvard.edu/abs/1972ApJ...176....1G}
}

@ARTICLE{Dressler1980,
       author = {{Dressler}, A.},
        title = "{Galaxy morphology in rich clusters: implications for the formation and evolution of galaxies.}",
      journal = {\apj},
         year = 1980,
        month = mar,
       volume = {236},
        pages = {351-365},
          doi = {10.1086/157753},
       adsurl = {https://ui.adsabs.harvard.edu/abs/1980ApJ...236..351D}
}

@ARTICLE{Moore1996,
       author = {{Moore}, Ben and {Katz}, Neal and {Lake}, George and {Dressler}, Alan and {Oemler}, Augustus},
        title = "{Galaxy harassment and the evolution of clusters of galaxies}",
      journal = {\nat},
         year = 1996,
        month = feb,
       volume = {379},
       number = {6566},
        pages = {613-616},
          doi = {10.1038/379613a0},
archivePrefix = {arXiv},
       eprint = {astro-ph/9510034},
 primaryClass = {astro-ph},
       adsurl = {https://ui.adsabs.harvard.edu/abs/1996Natur.379..613M}
}

@ARTICLE{Boselli2022,
       author = {{Boselli}, Alessandro and {Fossati}, Matteo and {Sun}, Ming},
        title = "{Ram pressure stripping in high-density environments}",
      journal = {\aapr},
         year = 2022,
        month = dec,
       volume = {30},
       number = {1},
          eid = {3},
        pages = {3},
          doi = {10.1007/s00159-022-00140-3},
archivePrefix = {arXiv},
       eprint = {2109.13614},
 primaryClass = {astro-ph.GA},
       adsurl = {https://ui.adsabs.harvard.edu/abs/2022A&ARv..30....3B}
}

@ARTICLE{Blanton2009,
       author = {{Blanton}, Michael R. and {Moustakas}, John},
        title = "{Physical Properties and Environments of Nearby Galaxies}",
      journal = {\araa},
         year = 2009,
        month = sep,
       volume = {47},
       number = {1},
        pages = {159-210},
          doi = {10.1146/annurev-astro-082708-101734},
archivePrefix = {arXiv},
       eprint = {0908.3017},
 primaryClass = {astro-ph.GA},
       adsurl = {https://ui.adsabs.harvard.edu/abs/2009ARA&A..47..159B}
}

@ARTICLE{Bahe2017,
       author = {{Bah{\'e}}, Yannick M. and {Barnes}, David J. and {Dalla Vecchia}, Claudio and {Kay}, Scott T. and {White}, Simon D.~M. and {McCarthy}, Ian G. and {Schaye}, Joop and {Bower}, Richard G. and {Crain}, Robert A. and {Theuns}, Tom and {Jenkins}, Adrian and {McGee}, Sean L. and {Schaller}, Matthieu and {Thomas}, Peter A. and {Trayford}, James W.},
        title = "{The Hydrangea simulations: galaxy formation in and around massive clusters}",
      journal = {\mnras},
         year = 2017,
        month = oct,
       volume = {470},
       number = {4},
        pages = {4186-4208},
          doi = {10.1093/mnras/stx1403},
archivePrefix = {arXiv},
       eprint = {1703.10610},
 primaryClass = {astro-ph.GA},
       adsurl = {https://ui.adsabs.harvard.edu/abs/2017MNRAS.470.4186B}
}

@ARTICLE{Jaffe2015,
       author = {{Jaff{\'e}}, Yara L. and {Smith}, Rory and {Candlish}, Graeme N. and {Poggianti}, Bianca M. and {Sheen}, Yun-Kyeong and {Verheijen}, Marc A.~W.},
        title = "{BUDHIES II: a phase-space view of H I gas stripping and star formation quenching in cluster galaxies}",
      journal = {\mnras},
         year = 2015,
        month = apr,
       volume = {448},
       number = {2},
        pages = {1715-1728},
          doi = {10.1093/mnras/stv100},
archivePrefix = {arXiv},
       eprint = {1501.03819},
 primaryClass = {astro-ph.GA},
       adsurl = {https://ui.adsabs.harvard.edu/abs/2015MNRAS.448.1715J}
}

@ARTICLE{Rhee2017,
       author = {{Rhee}, Jinsu and {Smith}, Rory and {Choi}, Hoseung and {Yi}, Sukyoung K. and {Jaff{\'e}}, Yara and {Candlish}, Graeme and {S{\'a}nchez-J{\'a}nssen}, Ruben},
        title = "{Phase-space Analysis in the Group and Cluster Environment: Time Since Infall and Tidal Mass Loss}",
      journal = {\apj},
         year = 2017,
        month = jul,
       volume = {843},
       number = {2},
          eid = {128},
        pages = {128},
          doi = {10.3847/1538-4357/aa6d6c},
archivePrefix = {arXiv},
       eprint = {1704.04243},
 primaryClass = {astro-ph.GA},
       adsurl = {https://ui.adsabs.harvard.edu/abs/2017ApJ...843..128R}
}

@ARTICLE{Croom2012,
       author = {{Croom}, Scott M. and {Lawrence}, Jon S. and {Bland-Hawthorn}, Joss and {Bryant}, Julia J. and {Fogarty}, Lisa and {Richards}, Samuel and {Goodwin}, Michael and {Farrell}, Tony and {Miziarski}, Stan and {Heald}, Ron and {Jones}, D. Heath and {Lee}, Steve and {Colless}, Matthew and {Brough}, Sarah and {Hopkins}, Andrew M. and {Bauer}, Amanda E. and {Birchall}, Michael N. and {Ellis}, Simon and {Horton}, Anthony and {Leon-Saval}, Sergio and {Lewis}, Geraint and {L{\'o}pez-S{\'a}nchez}, {\'A}. R. and {Min}, Seong-Sik and {Trinh}, Christopher and {Trowland}, Holly},
        title = "{The Sydney-AAO Multi-object Integral field spectrograph}",
      journal = {\mnras},
         year = 2012,
        month = mar,
       volume = {421},
       number = {1},
        pages = {872-893},
          doi = {10.1111/j.1365-2966.2011.20365.x},
archivePrefix = {arXiv},
       eprint = {1112.3367},
 primaryClass = {astro-ph.CO},
       adsurl = {https://ui.adsabs.harvard.edu/abs/2012MNRAS.421..872C}
}

@ARTICLE{Bundy2015,
       author = {{Bundy}, Kevin and {Bershady}, Matthew A. and {Law}, David R. and {Yan}, Renbin and {Drory}, Niv and {MacDonald}, Nicholas and {Wake}, David A. and {Cherinka}, Brian and {S{\'a}nchez-Gallego}, Jos{\'e} R. and {Weijmans}, Anne-Marie and {Thomas}, Daniel and {Tremonti}, Christy and {Masters}, Karen and {Coccato}, Lodovico and {Diamond-Stanic}, Aleksandar M. and {Arag{\'o}n-Salamanca}, Alfonso and {Avila-Reese}, Vladimir and {Badenes}, Carles and {Falc{\'o}n-Barroso}, J{\'e}sus and {Belfiore}, Francesco and {Bizyaev}, Dmitry and {Blanc}, Guillermo A. and {Bland-Hawthorn}, Joss and {Blanton}, Michael R. and {Brownstein}, Joel R. and {Byler}, Nell and {Cappellari}, Michele and {Conroy}, Charlie and {Dutton}, Aaron A. and {Emsellem}, Eric and {Etherington}, James and {Frinchaboy}, Peter M. and {Fu}, Hai and {Gunn}, James E. and {Harding}, Paul and {Johnston}, Evelyn J. and {Kauffmann}, Guinevere and {Kinemuchi}, Karen and {Klaene}, Mark A. and {Knapen}, Johan H. and {Leauthaud}, Alexie and {Li}, Cheng and {Lin}, Lihwai and {Maiolino}, Roberto and {Malanushenko}, Viktor and {Malanushenko}, Elena and {Mao}, Shude and {Maraston}, Claudia and {McDermid}, Richard M. and {Merrifield}, Michael R. and {Nichol}, Robert C. and {Oravetz}, Daniel and {Pan}, Kaike and {Parejko}, John K. and {Sanchez}, Sebastian F. and {Schlegel}, David and {Simmons}, Audrey and {Steele}, Oliver and {Steinmetz}, Matthias and {Thanjavur}, Karun and {Thompson}, Benjamin A. and {Tinker}, Jeremy L. and {van den Bosch}, Remco C.~E. and {Westfall}, Kyle B. and {Wilkinson}, David and {Wright}, Shelley and {Xiao}, Ting and {Zhang}, Kai},
        title = "{Overview of the SDSS-IV MaNGA Survey: Mapping nearby Galaxies at Apache Point Observatory}",
      journal = {\apj},
         year = 2015,
        month = jan,
       volume = {798},
       number = {1},
          eid = {7},
        pages = {7},
          doi = {10.1088/0004-637X/798/1/7},
archivePrefix = {arXiv},
       eprint = {1412.1482},
 primaryClass = {astro-ph.GA},
       adsurl = {https://ui.adsabs.harvard.edu/abs/2015ApJ...798....7B}
}

@ARTICLE{vandeSande2021,
       author = {{van de Sande}, Jesse and {Croom}, Scott M. and {Bland-Hawthorn}, Joss and {Cortese}, Luca and {Scott}, Nicholas and {Lagos}, Claudia D.~P. and {D'Eugenio}, Francesco and {Bryant}, Julia J. and {Brough}, Sarah and {Catinella}, Barbara and {Foster}, Caroline and {Groves}, Brent and {Harborne}, Katherine E. and {L{\'o}pez-S{\'a}nchez}, {\'A}ngel R. and {McDermid}, Richard and {Medling}, Anne and {Owers}, Matt S. and {Richards}, Samuel N. and {Sweet}, Sarah M. and {Vaughan}, Sam P.},
        title = "{The SAMI galaxy survey: Mass and environment as independent drivers of galaxy dynamics}",
      journal = {\mnras},
         year = 2021,
        month = dec,
       volume = {508},
       number = {2},
        pages = {2307-2328},
          doi = {10.1093/mnras/stab2647},
archivePrefix = {arXiv},
       eprint = {2109.06189},
 primaryClass = {astro-ph.GA},
       adsurl = {https://ui.adsabs.harvard.edu/abs/2021MNRAS.508.2307V}
}

@ARTICLE{Santucci2023,
       author = {{Santucci}, Giulia and {Brough}, Sarah and {van de Sande}, Jesse and {McDermid}, Richard and {Barsanti}, Stefania and {Bland-Hawthorn}, Joss and {Bryant}, Julia J. and {Croom}, Scott M. and {Lagos}, Claudia and {Lawrence}, Jon S. and {Owers}, Matt S. and {van de Ven}, Glenn and {Vaughan}, Sam P. and {Yi}, Sukyoung K.},
        title = "{The SAMI Galaxy Survey: Environmental analysis of the orbital structures of passive galaxies}",
      journal = {\mnras},
         year = 2023,
        month = may,
       volume = {521},
       number = {2},
        pages = {2671-2691},
          doi = {10.1093/mnras/stad713},
archivePrefix = {arXiv},
       eprint = {2303.04161},
 primaryClass = {astro-ph.GA},
       adsurl = {https://ui.adsabs.harvard.edu/abs/2023MNRAS.521.2671S}
}

@ARTICLE{Joshi2020,
       author = {{Joshi}, Gandhali D. and {Pillepich}, Annalisa and {Nelson}, Dylan and {Marinacci}, Federico and {Springel}, Volker and {Rodriguez-Gomez}, Vicente and {Vogelsberger}, Mark and {Hernquist}, Lars},
        title = "{The fate of disc galaxies in IllustrisTNG clusters}",
      journal = {\mnras},
         year = 2020,
        month = aug,
       volume = {496},
       number = {3},
        pages = {2673-2703},
          doi = {10.1093/mnras/staa1668},
archivePrefix = {arXiv},
       eprint = {2004.01191},
 primaryClass = {astro-ph.GA},
       adsurl = {https://ui.adsabs.harvard.edu/abs/2020MNRAS.496.2673J}
}

@ARTICLE{Lokas2020,
       author = {{{\L}okas}, Ewa L.},
        title = "{Tidal evolution of galaxies in the most massive cluster of IllustrisTNG-100}",
      journal = {\aap},
         year = 2020,
        month = jun,
       volume = {638},
          eid = {A133},
        pages = {A133},
          doi = {10.1051/0004-6361/202037643},
archivePrefix = {arXiv},
       eprint = {2002.00610},
 primaryClass = {astro-ph.GA},
       adsurl = {https://ui.adsabs.harvard.edu/abs/2020A&A...638A.133L}
}

@ARTICLE{Patton2020,
       author = {{Patton}, David R. and {Wilson}, Kieran D. and {Metrow}, Colin J. and {Ellison}, Sara L. and {Torrey}, Paul and {Brown}, Westley and {Hani}, Maan H. and {McAlpine}, Stuart and {Moreno}, Jorge and {Woo}, Joanna},
        title = "{Interacting galaxies in the IllustrisTNG simulations - I: Triggered star formation in a cosmological context}",
      journal = {\mnras},
         year = 2020,
        month = jun,
       volume = {494},
       number = {4},
        pages = {4969-4985},
          doi = {10.1093/mnras/staa913},
archivePrefix = {arXiv},
       eprint = {2003.00289},
 primaryClass = {astro-ph.GA},
       adsurl = {https://ui.adsabs.harvard.edu/abs/2020MNRAS.494.4969P}
}

@ARTICLE{Dou2025,
       author = {{Dou}, Haoran and {Yu}, Heng},
        title = "{Estimating infall times of galaxies around clusters: Working to achieve accurate results based on observational data}",
      journal = {\aap},
         year = 2025,
        month = jul,
       volume = {699},
          eid = {A95},
        pages = {A95},
          doi = {10.1051/0004-6361/202555240},
archivePrefix = {arXiv},
       eprint = {2505.17775},
 primaryClass = {astro-ph.GA},
       adsurl = {https://ui.adsabs.harvard.edu/abs/2025A&A...699A..95D}
}

@ARTICLE{Yun2019,
       author = {{Yun}, Kiyun and {Pillepich}, Annalisa and {Zinger}, Elad and {Nelson}, Dylan and {Donnari}, Martina and {Joshi}, Gandhali and {Rodriguez-Gomez}, Vicente and {Genel}, Shy and {Weinberger}, Rainer and {Vogelsberger}, Mark and {Hernquist}, Lars},
        title = "{Jellyfish galaxies with the IllustrisTNG simulations - I. Gas-stripping phenomena in the full cosmological context}",
      journal = {\mnras},
         year = 2019,
        month = feb,
       volume = {483},
       number = {1},
        pages = {1042-1066},
          doi = {10.1093/mnras/sty3156},
archivePrefix = {arXiv},
       eprint = {1810.00005},
 primaryClass = {astro-ph.GA},
       adsurl = {https://ui.adsabs.harvard.edu/abs/2019MNRAS.483.1042Y}
}

@ARTICLE{Rohr2023,
       author = {{Rohr}, Eric and {Pillepich}, Annalisa and {Nelson}, Dylan and {Zinger}, Elad and {Joshi}, Gandhali D. and {Ayromlou}, Mohammadreza},
        title = "{Jellyfish galaxies with the IllustrisTNG simulations - when, where, and for how long does ram pressure stripping of cold gas occur?}",
      journal = {\mnras},
         year = 2023,
        month = sep,
       volume = {524},
       number = {3},
        pages = {3502-3525},
          doi = {10.1093/mnras/stad2101},
archivePrefix = {arXiv},
       eprint = {2304.09196},
 primaryClass = {astro-ph.GA},
       adsurl = {https://ui.adsabs.harvard.edu/abs/2023MNRAS.524.3502R}
}

@ARTICLE{Goller2023,
       author = {{G{\"o}ller}, Junia and {Joshi}, Gandhali D. and {Rohr}, Eric and {Zinger}, Elad and {Pillepich}, Annalisa},
        title = "{Jellyfish galaxies with the IllustrisTNG simulations - No enhanced population-wide star formation according to TNG50}",
      journal = {\mnras},
         year = 2023,
        month = nov,
       volume = {525},
       number = {3},
        pages = {3551-3570},
          doi = {10.1093/mnras/stad2551},
archivePrefix = {arXiv},
       eprint = {2304.09199},
 primaryClass = {astro-ph.GA},
       adsurl = {https://ui.adsabs.harvard.edu/abs/2023MNRAS.525.3551G}
}

@ARTICLE{Zinger2024,
       author = {{Zinger}, Elad and {Joshi}, Gandhali D. and {Pillepich}, Annalisa and {Rohr}, Eric and {Nelson}, Dylan},
        title = "{Jellyfish galaxies with the IllustrisTNG simulations - citizen-science results towards large distances, low-mass hosts, and high redshifts}",
      journal = {\mnras},
         year = 2024,
        month = jan,
       volume = {527},
       number = {3},
        pages = {8257-8289},
          doi = {10.1093/mnras/stad3716},
archivePrefix = {arXiv},
       eprint = {2304.09202},
 primaryClass = {astro-ph.GA},
       adsurl = {https://ui.adsabs.harvard.edu/abs/2024MNRAS.527.8257Z}
}

@ARTICLE{KurinchiVendhan2025,
       author = {{Kurinchi-Vendhan}, Shalini and {Rohr}, Eric and {Pillepich}, Annalisa and {Zinger}, Elad and {Ayromlou}, Mohammadreza and {Joshi}, Gandhali D.},
        title = "{Jellyfish galaxies with the IllustrisTNG simulations {\textendash} Supermassive black hole activity in dense environments with ram-pressure stripped satellites}",
      journal = {\mnras},
         year = 2025,
        month = sep,
       volume = {542},
       number = {3},
        pages = {1901-1922},
          doi = {10.1093/mnras/staf1280},
archivePrefix = {arXiv},
       eprint = {2506.05474},
 primaryClass = {astro-ph.GA},
       adsurl = {https://ui.adsabs.harvard.edu/abs/2025MNRAS.542.1901K}
}

@ARTICLE{Kang2008,
       author = {{Kang}, X. and {van den Bosch}, Frank C.},
        title = "{New Constraints on the Efficiencies of Ram Pressure Stripping and the Tidal Disruption of Satellite Galaxies}",
      journal = {\apjl},
         year = 2008,
        month = apr,
       volume = {676},
       number = {2},
        pages = {L101},
          doi = {10.1086/587620},
archivePrefix = {arXiv},
       eprint = {0801.1843},
 primaryClass = {astro-ph},
       adsurl = {https://ui.adsabs.harvard.edu/abs/2008ApJ...676L.101K}
}

@ARTICLE{Vulcani2021,
       author = {{Vulcani}, Benedetta and {Poggianti}, Bianca M. and {Moretti}, Alessia and {Franchetto}, Andrea and {Bacchini}, Cecilia and {McGee}, Sean and {Jaff{\'e}}, Yara L. and {Mingozzi}, Matilde and {Werle}, Ariel and {Tomi{\v{c}}i{\'c}}, Neven and {Fritz}, Jacopo and {Bettoni}, Daniela and {Wolter}, Anna and {Gullieuszik}, Marco},
        title = "{GASP. XXXIII. The Ability of Spatially Resolved Data to Distinguish among the Different Physical Mechanisms Affecting Galaxies in Low-density Environments}",
      journal = {\apj},
         year = 2021,
        month = jun,
       volume = {914},
       number = {1},
          eid = {27},
        pages = {27},
          doi = {10.3847/1538-4357/abf655},
archivePrefix = {arXiv},
       eprint = {2104.02089},
 primaryClass = {astro-ph.GA},
       adsurl = {https://ui.adsabs.harvard.edu/abs/2021ApJ...914...27V}
}

@ARTICLE{Larson1980,
       author = {{Larson}, R.~B. and {Tinsley}, B.~M. and {Caldwell}, C.~N.},
        title = "{The evolution of disk galaxies and the origin of S0 galaxies}",
      journal = {\apj},
         year = 1980,
        month = may,
       volume = {237},
        pages = {692-707},
          doi = {10.1086/157917},
       adsurl = {https://ui.adsabs.harvard.edu/abs/1980ApJ...237..692L}
}

@ARTICLE{Trussler2020,
       author = {{Trussler}, James and {Maiolino}, Roberto and {Maraston}, Claudia and {Peng}, Yingjie and {Thomas}, Daniel and {Goddard}, Daniel and {Lian}, Jianhui},
        title = "{Both starvation and outflows drive galaxy quenching}",
      journal = {\mnras},
         year = 2020,
        month = feb,
       volume = {491},
       number = {4},
        pages = {5406-5434},
          doi = {10.1093/mnras/stz3286},
archivePrefix = {arXiv},
       eprint = {1811.09283},
 primaryClass = {astro-ph.GA},
       adsurl = {https://ui.adsabs.harvard.edu/abs/2020MNRAS.491.5406T}
}

@ARTICLE{Roberts2021,
       author = {{Roberts}, I.~D. and {van Weeren}, R.~J. and {McGee}, S.~L. and {Botteon}, A. and {Drabent}, A. and {Ignesti}, A. and {Rottgering}, H.~J.~A. and {Shimwell}, T.~W. and {Tasse}, C.},
        title = "{LoTSS jellyfish galaxies. I. Radio tails in low redshift clusters}",
      journal = {\aap},
         year = 2021,
        month = jun,
       volume = {650},
          eid = {A111},
        pages = {A111},
          doi = {10.1051/0004-6361/202140784},
archivePrefix = {arXiv},
       eprint = {2104.05383},
 primaryClass = {astro-ph.GA},
       adsurl = {https://ui.adsabs.harvard.edu/abs/2021A&A...650A.111R}
}

@ARTICLE{Poggianti2017,
       author = {{Poggianti}, Bianca M. and {Moretti}, Alessia and {Gullieuszik}, Marco and {Fritz}, Jacopo and {Jaff{\'e}}, Yara and {Bettoni}, Daniela and {Fasano}, Giovanni and {Bellhouse}, Callum and {Hau}, George and {Vulcani}, Benedetta and {Biviano}, Andrea and {Omizzolo}, Alessandro and {Paccagnella}, Angela and {D'Onofrio}, Mauro and {Cava}, Antonio and {Sheen}, Y.-K. and {Couch}, Warrick and {Owers}, Matt},
        title = "{GASP. I. Gas Stripping Phenomena in Galaxies with MUSE}",
      journal = {\apj},
         year = 2017,
        month = jul,
       volume = {844},
       number = {1},
          eid = {48},
        pages = {48},
          doi = {10.3847/1538-4357/aa78ed},
archivePrefix = {arXiv},
       eprint = {1704.05086},
 primaryClass = {astro-ph.GA},
       adsurl = {https://ui.adsabs.harvard.edu/abs/2017ApJ...844...48P}
}

@ARTICLE{Gill2004,
       author = {{Gill}, Stuart P.~D. and {Knebe}, Alexander and {Gibson}, Brad K.},
        title = "{The evolution of substructure - I. A new identification method}",
      journal = {\mnras},
         year = 2004,
        month = jun,
       volume = {351},
       number = {2},
        pages = {399-409},
          doi = {10.1111/j.1365-2966.2004.07786.x},
archivePrefix = {arXiv},
       eprint = {astro-ph/0404258},
 primaryClass = {astro-ph},
       adsurl = {https://ui.adsabs.harvard.edu/abs/2004MNRAS.351..399G}
}

@ARTICLE{Gill2005,
       author = {{Gill}, Stuart P.~D. and {Knebe}, Alexander and {Gibson}, Brad K.},
        title = "{The evolution of substructure - III. The outskirts of clusters}",
      journal = {\mnras},
         year = 2005,
        month = feb,
       volume = {356},
       number = {4},
        pages = {1327-1332},
          doi = {10.1111/j.1365-2966.2004.08562.x},
archivePrefix = {arXiv},
       eprint = {astro-ph/0404427},
 primaryClass = {astro-ph},
       adsurl = {https://ui.adsabs.harvard.edu/abs/2005MNRAS.356.1327G}
}

@ARTICLE{Mahajan2011,
       author = {{Mahajan}, Smriti and {Mamon}, Gary A. and {Raychaudhury}, Somak},
        title = "{The velocity modulation of galaxy properties in and near clusters: quantifying the decrease in star formation in backsplash galaxies}",
      journal = {\mnras},
         year = 2011,
        month = oct,
       volume = {416},
       number = {4},
        pages = {2882-2902},
          doi = {10.1111/j.1365-2966.2011.19236.x},
archivePrefix = {arXiv},
       eprint = {1106.3062},
 primaryClass = {astro-ph.CO},
       adsurl = {https://ui.adsabs.harvard.edu/abs/2011MNRAS.416.2882M}
}

@ARTICLE{Pasquali2019,
       author = {{Pasquali}, A. and {Smith}, R. and {Gallazzi}, A. and {De Lucia}, G. and {Zibetti}, S. and {Hirschmann}, M. and {Yi}, S.~K.},
        title = "{Physical properties of SDSS satellite galaxies in projected phase space}",
      journal = {\mnras},
         year = 2019,
        month = apr,
       volume = {484},
       number = {2},
        pages = {1702-1723},
          doi = {10.1093/mnras/sty3530},
archivePrefix = {arXiv},
       eprint = {1901.04238},
 primaryClass = {astro-ph.GA},
       adsurl = {https://ui.adsabs.harvard.edu/abs/2019MNRAS.484.1702P}
}

@ARTICLE{Biviano2013,
       author = {{Biviano}, A. and {Rosati}, P. and {Balestra}, I. and {Mercurio}, A. and {Girardi}, M. and {Nonino}, M. and {Grillo}, C. and {Scodeggio}, M. and {Lemze}, D. and {Kelson}, D. and {Umetsu}, K. and {Postman}, M. and {Zitrin}, A. and {Czoske}, O. and {Ettori}, S. and {Fritz}, A. and {Lombardi}, M. and {Maier}, C. and {Medezinski}, E. and {Mei}, S. and {Presotto}, V. and {Strazzullo}, V. and {Tozzi}, P. and {Ziegler}, B. and {Annunziatella}, M. and {Bartelmann}, M. and {Benitez}, N. and {Bradley}, L. and {Brescia}, M. and {Broadhurst}, T. and {Coe}, D. and {Demarco}, R. and {Donahue}, M. and {Ford}, H. and {Gobat}, R. and {Graves}, G. and {Koekemoer}, A. and {Kuchner}, U. and {Melchior}, P. and {Meneghetti}, M. and {Merten}, J. and {Moustakas}, L. and {Munari}, E. and {Reg{\H{o}}s}, E. and {Sartoris}, B. and {Seitz}, S. and {Zheng}, W.},
        title = "{CLASH-VLT: The mass, velocity-anisotropy, and pseudo-phase-space density profiles of the z = 0.44 galaxy cluster MACS J1206.2-0847}",
      journal = {\aap},
         year = 2013,
        month = oct,
       volume = {558},
          eid = {A1},
        pages = {A1},
          doi = {10.1051/0004-6361/201321955},
archivePrefix = {arXiv},
       eprint = {1307.5867},
 primaryClass = {astro-ph.CO},
       adsurl = {https://ui.adsabs.harvard.edu/abs/2013A&A...558A...1B}
}

@ARTICLE{Fogarty2014,
       author = {{Fogarty}, L.~M.~R. and {Scott}, Nicholas and {Owers}, Matt S. and {Brough}, S. and {Croom}, Scott M. and {Pracy}, Michael B. and {Houghton}, R.~C.~W. and {Bland-Hawthorn}, Joss and {Colless}, Matthew and {Davies}, Roger L. and {Jones}, D. Heath and {Allen}, J.~T. and {Bryant}, Julia J. and {Goodwin}, Michael and {Green}, Andrew W. and {Konstantopoulos}, Iraklis S. and {Lawrence}, J.~S. and {Richards}, Samuel and {Cortese}, Luca and {Sharp}, Rob},
        title = "{The SAMI Pilot Survey: the kinematic morphology-density relation in Abell 85, Abell 168 and Abell 2399}",
      journal = {\mnras},
         year = 2014,
        month = sep,
       volume = {443},
       number = {1},
        pages = {485-503},
          doi = {10.1093/mnras/stu1165},
archivePrefix = {arXiv},
       eprint = {1406.3899},
 primaryClass = {astro-ph.GA},
       adsurl = {https://ui.adsabs.harvard.edu/abs/2014MNRAS.443..485F}
}

@ARTICLE{Fogarty2015,
       author = {{Fogarty}, L.~M.~R. and {Scott}, N. and {Owers}, M.~S. and {Croom}, S.~M. and {Bekki}, K. and {Houghton}, R.~C.~W. and {van de Sande}, J. and {D'Eugenio}, F. and {Cecil}, G.~N. and {Colless}, M.~M. and {Bland-Hawthorn}, J. and {Brough}, S. and {Cortese}, L. and {Davies}, R.~L. and {Jones}, D.~H. and {Pracy}, M. and {Allen}, J.~T. and {Bryant}, J.~J. and {Goodwin}, M. and {Green}, A.~W. and {Konstantopoulos}, I.~S. and {Lawrence}, J.~S. and {Lorente}, N.~P.~F. and {Richards}, S. and {Sharp}, R.~G.},
        title = "{The SAMI Pilot Survey: stellar kinematics of galaxies in Abell 85, 168 and 2399}",
      journal = {\mnras},
         year = 2015,
        month = dec,
       volume = {454},
       number = {2},
        pages = {2050-2066},
          doi = {10.1093/mnras/stv2060},
archivePrefix = {arXiv},
       eprint = {1509.02641},
 primaryClass = {astro-ph.GA},
       adsurl = {https://ui.adsabs.harvard.edu/abs/2015MNRAS.454.2050F}
}

@ARTICLE{Brough2017,
       author = {{Brough}, Sarah and {van de Sande}, Jesse and {Owers}, Matt S. and {d'Eugenio}, Francesco and {Sharp}, Rob and {Cortese}, Luca and {Scott}, Nicholas and {Croom}, Scott M. and {Bassett}, Rob and {Bekki}, Kenji and {Bland-Hawthorn}, Joss and {Bryant}, Julia J. and {Davies}, Roger and {Drinkwater}, Michael J. and {Driver}, Simon P. and {Foster}, Caroline and {Goldstein}, Gregory and {L{\'o}pez-S{\'a}nchez}, {\'A}. R. and {Medling}, Anne M. and {Sweet}, Sarah M. and {Taranu}, Dan S. and {Tonini}, Chiara and {Yi}, Sukyoung K. and {Goodwin}, Michael and {Lawrence}, J.~S. and {Richards}, Samuel N.},
        title = "{The SAMI Galaxy Survey: Mass as the Driver of the Kinematic Morphology-Density Relation in Clusters}",
      journal = {\apj},
         year = 2017,
        month = jul,
       volume = {844},
       number = {1},
          eid = {59},
        pages = {59},
          doi = {10.3847/1538-4357/aa7a11},
archivePrefix = {arXiv},
       eprint = {1704.01169},
 primaryClass = {astro-ph.GA},
       adsurl = {https://ui.adsabs.harvard.edu/abs/2017ApJ...844...59B}
}

@ARTICLE{Cortese2019,
       author = {{Cortese}, L. and {van de Sande}, J. and {Lagos}, C.~P. and {Catinella}, B. and {Davies}, L.~J.~M. and {Croom}, S.~M. and {Brough}, S. and {Bryant}, J.~J. and {Lawrence}, J.~S. and {Owers}, M.~S. and {Richards}, S.~N. and {Sweet}, S.~M. and {Bland-Hawthorn}, J.},
        title = "{The SAMI Galaxy Survey: satellite galaxies undergo little structural change during their quenching phase}",
      journal = {\mnras},
         year = 2019,
        month = may,
       volume = {485},
       number = {2},
        pages = {2656-2665},
          doi = {10.1093/mnras/stz485},
archivePrefix = {arXiv},
       eprint = {1902.05652},
 primaryClass = {astro-ph.GA},
       adsurl = {https://ui.adsabs.harvard.edu/abs/2019MNRAS.485.2656C}
}

@ARTICLE{Barsanti2025,
       author = {{Barsanti}, Stefania and {Croom}, Scott M. and {Colless}, Matthew and {Bland-Hawthorn}, Joss and {Brough}, Sarah and {Bryant}, Julia J. and {Lorente}, Nuria and {Oh}, Sree and {Santucci}, Giulia and {Sweet}, Sarah and {Sande van de}, Jesse and {Welker}, Charlotte},
        title = "{The SAMI Galaxy Survey: large-scale environment affects galaxy spin amplitudes and the formation of slow rotators}",
      journal = {\mnras},
         year = 2025,
        month = apr,
       volume = {538},
       number = {4},
        pages = {2660-2675},
          doi = {10.1093/mnras/staf426},
archivePrefix = {arXiv},
       eprint = {2503.09052},
 primaryClass = {astro-ph.GA},
       adsurl = {https://ui.adsabs.harvard.edu/abs/2025MNRAS.538.2660B}
}

@ARTICLE{Medling2018,
       author = {{Medling}, Anne M. and {Cortese}, Luca and {Croom}, Scott M. and {Green}, Andrew W. and {Groves}, Brent and {Hampton}, Elise and {Ho}, I.-Ting and {Davies}, Luke J.~M. and {Kewley}, Lisa J. and {Moffett}, Amanda J. and {Schaefer}, Adam L. and {Taylor}, Edward and {Zafar}, Tayyaba and {Bekki}, Kenji and {Bland-Hawthorn}, Joss and {Bloom}, Jessica V. and {Brough}, Sarah and {Bryant}, Julia J. and {Catinella}, Barbara and {Cecil}, Gerald and {Colless}, Matthew and {Couch}, Warrick J. and {Drinkwater}, Michael J. and {Driver}, Simon P. and {Federrath}, Christoph and {Foster}, Caroline and {Goldstein}, Gregory and {Goodwin}, Michael and {Hopkins}, Andrew and {Lawrence}, J.~S. and {Leslie}, Sarah K. and {Lewis}, Geraint F. and {Lorente}, Nuria P.~F. and {Owers}, Matt S. and {McDermid}, Richard and {Richards}, Samuel N. and {Sharp}, Robert and {Scott}, Nicholas and {Sweet}, Sarah M. and {Taranu}, Dan S. and {Tescari}, Edoardo and {Tonini}, Chiara and {van de Sande}, Jesse and {Walcher}, C. Jakob and {Wright}, Angus},
        title = "{The SAMI Galaxy Survey: spatially resolving the main sequence of star formation}",
      journal = {\mnras},
         year = 2018,
        month = apr,
       volume = {475},
       number = {4},
        pages = {5194-5214},
          doi = {10.1093/mnras/sty127},
archivePrefix = {arXiv},
       eprint = {1801.04283},
 primaryClass = {astro-ph.GA},
       adsurl = {https://ui.adsabs.harvard.edu/abs/2018MNRAS.475.5194M}
}

@ARTICLE{Dolag2009,
       author = {{Dolag}, K. and {Borgani}, S. and {Murante}, G. and {Springel}, V.},
        title = "{Substructures in hydrodynamical cluster simulations}",
      journal = {\mnras},
         year = 2009,
        month = oct,
       volume = {399},
       number = {2},
        pages = {497-514},
          doi = {10.1111/j.1365-2966.2009.15034.x},
archivePrefix = {arXiv},
       eprint = {0808.3401},
 primaryClass = {astro-ph},
       adsurl = {https://ui.adsabs.harvard.edu/abs/2009MNRAS.399..497D}
}

@ARTICLE{Donnari2019,
       author = {{Donnari}, Martina and {Pillepich}, Annalisa and {Nelson}, Dylan and {Vogelsberger}, Mark and {Genel}, Shy and {Weinberger}, Rainer and {Marinacci}, Federico and {Springel}, Volker and {Hernquist}, Lars},
        title = "{The star formation activity of IllustrisTNG galaxies: main sequence, UVJ diagram, quenched fractions, and systematics}",
      journal = {\mnras},
         year = 2019,
        month = jun,
       volume = {485},
       number = {4},
        pages = {4817-4840},
          doi = {10.1093/mnras/stz712},
archivePrefix = {arXiv},
       eprint = {1812.07584},
 primaryClass = {astro-ph.GA},
       adsurl = {https://ui.adsabs.harvard.edu/abs/2019MNRAS.485.4817D}
}

@ARTICLE{Davies2019,
       author = {{Davies}, L.~J.~M. and {Lagos}, C. del P. and {Katsianis}, A. and {Robotham}, A.~S.~G. and {Cortese}, L. and {Driver}, S.~P. and {Bremer}, M.~N. and {Brown}, M.~J.~I. and {Brough}, S. and {Cluver}, M.~E. and {Grootes}, M.~W. and {Holwerda}, B.~W. and {Owers}, M. and {Phillipps}, S.},
        title = "{Galaxy And Mass Assembly (GAMA): The sSFR-M$_{*}$ relation part I - {\ensuremath{\sigma}}$_{sSFR}$-M$_{*}$ as a function of sample, SFR indicator, and environment}",
      journal = {\mnras},
         year = 2019,
        month = feb,
       volume = {483},
       number = {2},
        pages = {1881-1900},
          doi = {10.1093/mnras/sty2957},
archivePrefix = {arXiv},
       eprint = {1811.03712},
 primaryClass = {astro-ph.GA},
       adsurl = {https://ui.adsabs.harvard.edu/abs/2019MNRAS.483.1881D}
}

@ARTICLE{Lokas2023,
       author = {{{\L}okas}, Ewa L.},
        title = "{Preprocessing in small groups: Three simulated galaxies interacting prior to cluster infall}",
      journal = {\aap},
         year = 2023,
        month = oct,
       volume = {678},
          eid = {A147},
        pages = {A147},
          doi = {10.1051/0004-6361/202347735},
archivePrefix = {arXiv},
       eprint = {2309.07494},
 primaryClass = {astro-ph.GA},
       adsurl = {https://ui.adsabs.harvard.edu/abs/2023A&A...678A.147L}
}

@ARTICLE{Masson2026,
       author = {{Masson}, Florine and {Parker}, Laura C.},
        title = "{Calibrating Galaxy Infall Times in Groups and Clusters with IllustrisTNG Simulations}",
      journal = {\apj},
         year = 2026,
        month = apr,
       volume = {1000},
       number = {2},
          eid = {194},
        pages = {194},
          doi = {10.3847/1538-4357/ae4e21},
archivePrefix = {arXiv},
       eprint = {2603.13010},
 primaryClass = {astro-ph.GA},
       adsurl = {https://ui.adsabs.harvard.edu/abs/2026ApJ..1000..194M}
}

@ARTICLE{Oman2013,
       author = {{Oman}, Kyle A. and {Hudson}, Michael J. and {Behroozi}, Peter S.},
        title = "{Disentangling satellite galaxy populations using orbit tracking in simulations}",
      journal = {\mnras},
         year = 2013,
        month = may,
       volume = {431},
       number = {3},
        pages = {2307-2316},
          doi = {10.1093/mnras/stt328},
archivePrefix = {arXiv},
       eprint = {1301.6757},
 primaryClass = {astro-ph.CO},
       adsurl = {https://ui.adsabs.harvard.edu/abs/2013MNRAS.431.2307O}
}

@ARTICLE{Oman2021,
       author = {{Oman}, Kyle A. and {Bah{\'e}}, Yannick M. and {Healy}, Julia and {Hess}, Kelley M. and {Hudson}, Michael J. and {Verheijen}, Marc A.~W.},
        title = "{A homogeneous measurement of the delay between the onsets of gas stripping and star formation quenching in satellite galaxies of groups and clusters}",
      journal = {\mnras},
         year = 2021,
        month = mar,
       volume = {501},
       number = {4},
        pages = {5073-5095},
          doi = {10.1093/mnras/staa3845},
archivePrefix = {arXiv},
       eprint = {2009.00667},
 primaryClass = {astro-ph.GA},
       adsurl = {https://ui.adsabs.harvard.edu/abs/2021MNRAS.501.5073O}
}

@ARTICLE{Rhee2020,
       author = {{Rhee}, Jinsu and {Smith}, Rory and {Choi}, Hoseung and {Contini}, Emanuele and {Jung}, S. Lyla and {Han}, San and {Yi}, Sukyoung K.},
        title = "{YZiCS: Unveiling the Quenching History of Cluster Galaxies Using Phase-space Analysis}",
      journal = {\apjs},
         year = 2020,
        month = apr,
       volume = {247},
       number = {2},
          eid = {45},
        pages = {45},
          doi = {10.3847/1538-4365/ab7377},
archivePrefix = {arXiv},
       eprint = {2002.04645},
 primaryClass = {astro-ph.GA},
       adsurl = {https://ui.adsabs.harvard.edu/abs/2020ApJS..247...45R}
}

@ARTICLE{Wetzel2013,
       author = {{Wetzel}, Andrew R. and {Tinker}, Jeremy L. and {Conroy}, Charlie and {van den Bosch}, Frank C.},
        title = "{Galaxy evolution in groups and clusters: satellite star formation histories and quenching time-scales in a hierarchical Universe}",
      journal = {\mnras},
         year = 2013,
        month = jun,
       volume = {432},
       number = {1},
        pages = {336-358},
          doi = {10.1093/mnras/stt469},
archivePrefix = {arXiv},
       eprint = {1206.3571},
 primaryClass = {astro-ph.CO},
       adsurl = {https://ui.adsabs.harvard.edu/abs/2013MNRAS.432..336W}
}

@ARTICLE{deIsidio2026,
       author = {{de Is{\'\i}dio}, Natan and {Popesso}, Paola and {Tacchella}, Sandro and {Pasquali}, Anna and {Marini}, Ilaria and {Mazengo}, Daudi and {Toptun}, Victoria and {McGee}, Sean},
        title = "{Circumgalactic medium depletion drives satellite quenching in IllustrisTNG}",
      journal = {arXiv e-prints},
         year = 2026,
        month = aug,
          eid = {arXiv:2608.03693},
        pages = {arXiv:2608.03693},
          doi = {10.48550/arXiv.2608.03693},
archivePrefix = {arXiv},
       eprint = {2608.03693},
 primaryClass = {astro-ph.GA},
       adsurl = {https://ui.adsabs.harvard.edu/abs/2026arXiv260803693D}
}

\clearpage
\appendix
\section{The simulated mass distribution}
\label{appendix A}

\begin{figure*}
\centering
\includegraphics{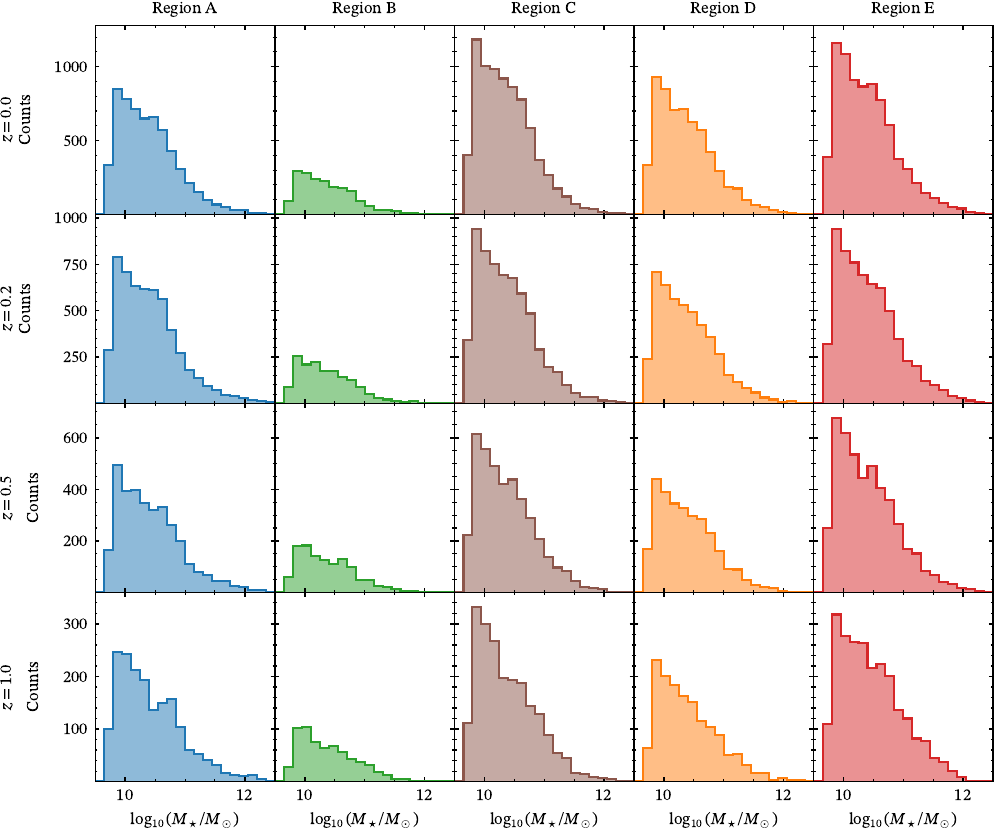}
\caption{Stellar-mass distributions across projected phase-space infall regions for the TNG-Cluster sample at four simulation snapshots. Columns correspond to regions A--E (as defined in the phase-space diagrams), and each panel shows the stellar-mass histogram of satellite galaxies assigned to the corresponding region. Rows indicate the four epochs analysed, at redshifts of $z = 0.0$, $0.2$, $0.5$, and $1.0$ (from top to bottom).}
\label{fig:figA1}
\end{figure*}

This appendix provides additional information about the sample composition in the projected phase-space regions analysed in this work. Table~\ref{tab:region_counts} lists the number of satellite galaxies and their corresponding fractions in each region (A--E) at the four redshifts considered in our analysis. Figure~\ref{fig:figA1} shows the stellar-mass distributions of the satellites in each region.

As described in Section~\ref{cluster sample selection} we restrict the analysis to galaxies with stellar masses $M_\star \geq 10^{9.75}\,\mathrm M_\odot$. This selection is motivated by the need to obtain reliable estimates of the stellar kinematic spin parameter. Although this mass limit reduces the number of satellites, all projected phase-space regions remain well populated across the snapshots considered.

However, the impact of this mass selection varies with redshift. Because the typical stellar masses of cluster satellites decrease at higher redshifts, the mass cut removes a larger fraction of galaxies at $z = 1.0$. As a result, the reduction in satellite counts is more significant at $z = 1.0$ than at $z = 0.0$, which should be kept in mind when interpreting evolutionary trends. The reliability of the kinematic measurements at this mass limit is examined in Appendix~\ref{appendix C}.

\begin{table*}
\centering
\caption{Number of satellite galaxies and corresponding fractions in each projected phase-space region at different redshifts. Fractions are given in parentheses (per cent).}
\label{tab:region_counts}
\begin{tabular}{c c c c c c}
\hline
$z$ & A & B & C & D & E \\
\hline
0.0 & 5971 (20.02) & 1906 (6.39) & 7841 (26.30) & 6080 (20.39) & 8021 (26.90) \\
0.2 & 5516 (22.47) & 1629 (6.64) & 6232 (25.39) & 4711 (19.19) & 6460 (26.32) \\
0.5 & 3328 (20.63) & 1195 (7.41) & 4029 (24.98) & 2950 (18.29) & 4627 (28.69) \\
1.0 & 1783 (21.19) & 649 (7.71)  & 2097 (24.92) & 1483 (17.62) & 2403 (28.56) \\
\hline
\end{tabular}
\end{table*}

\section{The simulated $\lambda_R$ distribution}
\label{appendix B}

\begin{figure*}
\centering
\includegraphics{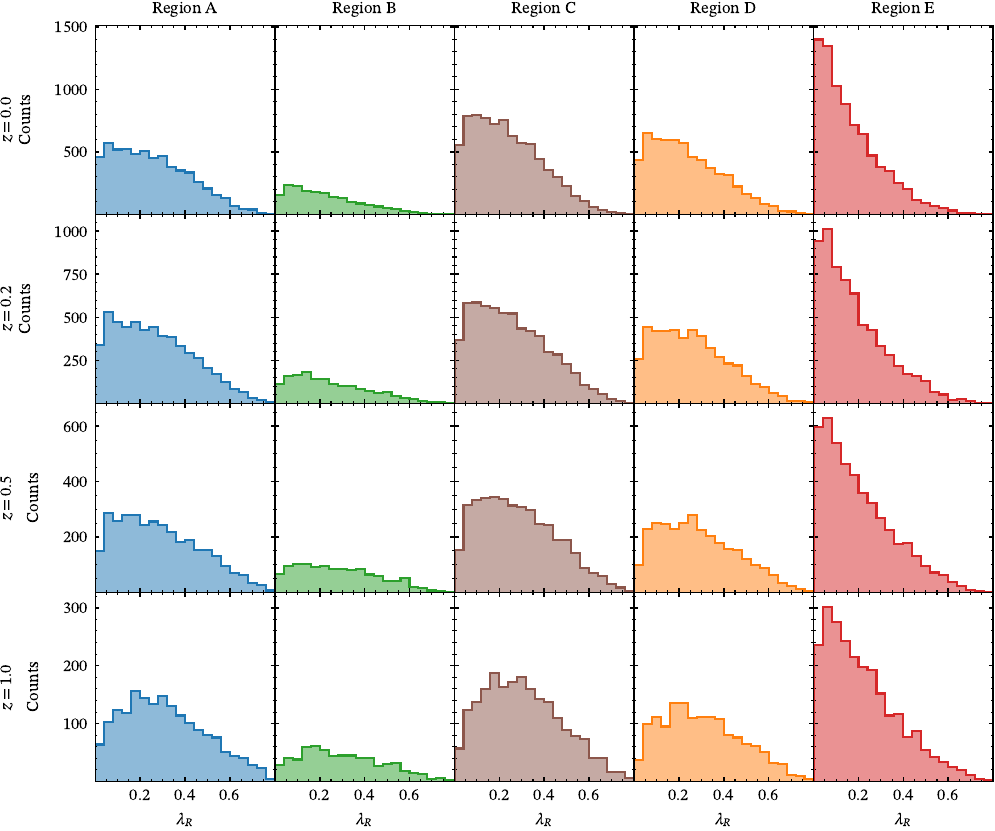}
\caption{$\lambda_R$ distributions across projected phase-space infall regions for the TNG-Cluster sample. Columns correspond to regions A--E, and each panel presents the $\lambda_R$ histogram of satellite galaxies assigned to that region. Rows indicate the four simulation snapshots analysed, at redshifts of $z = 0.0$, $0.2$, 
$0.5$, and $1.0$ (from top to bottom), highlighting the evolution of the $\lambda_R$ distribution within each infall region.}
\label{fig:figB1}
\end{figure*}

This appendix provides additional information about the spin-parameter distributions of satellite galaxies in the projected phase-space regions analysed in this work. Figure~\ref{fig:figB1} shows the distributions of the spin parameter $\lambda_R$ for satellites in each region (A--E) at the four redshifts considered.

While the median values are discussed in Section~\ref{evolution infall regions}, these distributions highlight the spread of $\lambda_R$ within each region and epoch. Galaxies in regions A--D generally exhibit similar distributions with relatively high spin, whereas satellites in region E display systematically lower spin values, consistent with the interpretation of this population as dynamically older and more virialised. The figure provides additional context for understanding the temporal evolution trends shown in Fig.~\ref{fig:fig5}.

\section{Resolution and the reliability of the simulated $\lambda_R$}
\label{appendix C}

This appendix presents additional tests supporting the discussion in Section~\ref{sec:caveats}. The mass resolution of TNG-Cluster ($m_{\rm DM} \approx 5.9 \times 10^7\,\mathrm{M_\odot}$, $m_{\rm gas} \approx 1.1 \times 10^7\,\mathrm{M_\odot}$) limits the recovery of internal stellar structure in individual satellites, particularly near the stellar-mass threshold adopted in this work. The simulated $\lambda_R$ values are therefore interpreted as a proxy for stellar rotational support rather than as measurements directly comparable, in absolute value, to the SAMI measurements.

\begin{figure*}
\centering
\includegraphics{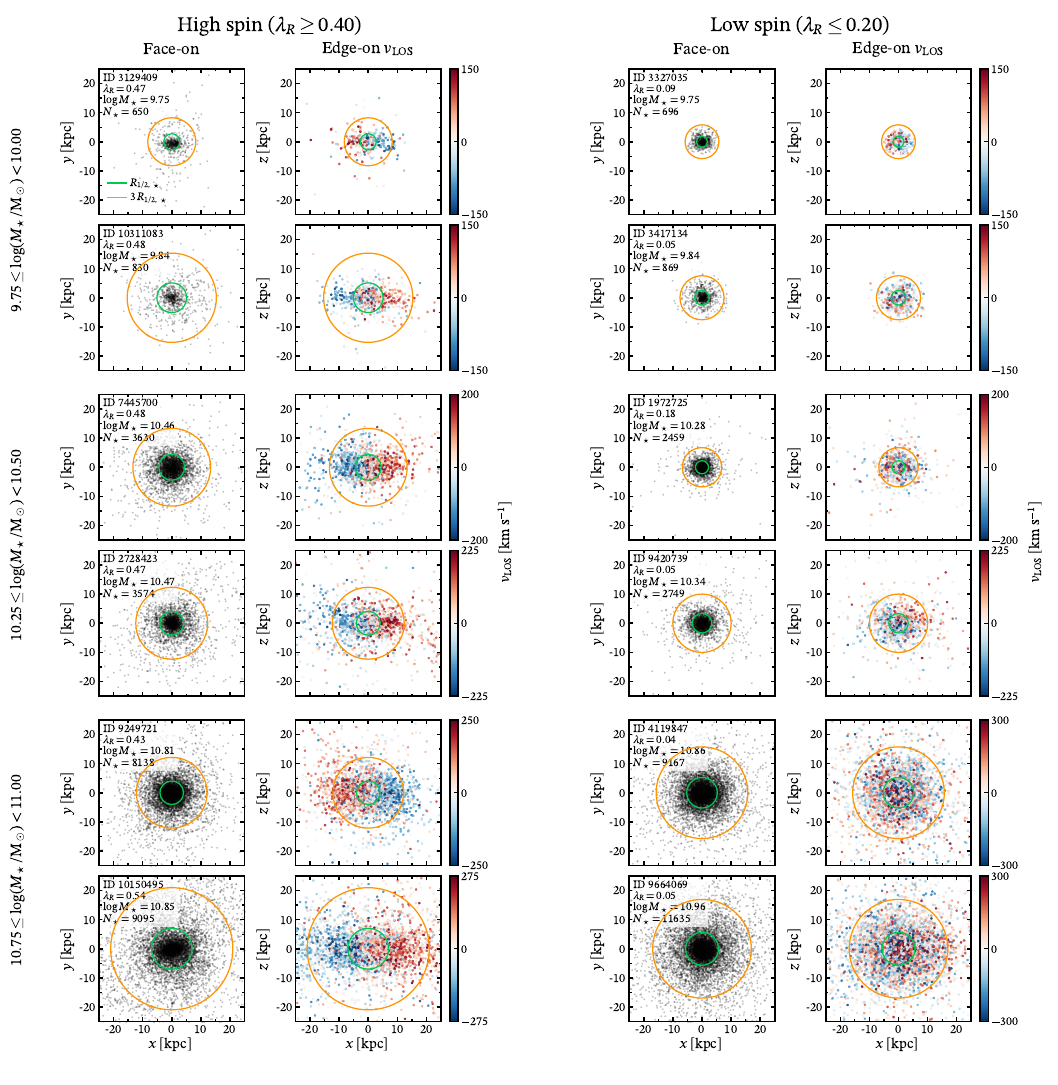}
\caption{Stellar particle distributions of twelve representative TNG-Cluster satellites at $z=0$, spanning the stellar-mass range adopted in this work. High-spin galaxies ($\lambda_R \geq 0.40$) are shown on the left and low-spin galaxies ($\lambda_R \leq 0.20$) on the right. For each galaxy, the face-on stellar distribution is shown next to the corresponding edge-on projection, where particles are colour-coded by the line-of-sight velocity $v_{\rm LOS}$. The velocity scale is set independently for each galaxy and is given by the corresponding colour bar. The sample is arranged in three stellar-mass bins, with two galaxies per bin. Green and orange circles indicate the stellar half-mass radius $R_{1/2,\star}$ and $3R_{1/2,\star}$, respectively; the latter is the aperture used to measure $\lambda_R$ (Section~\ref{spin parameter simulation}). The face-on panels also list the subhalo ID, $\lambda_R$, stellar mass, and number of bound stellar particles $N_\star$.}
\label{fig:figC1}
\end{figure*}

Figure~\ref{fig:figC1} shows the stellar particle distributions of twelve satellites spanning three stellar-mass bins and two spin regimes. Within each mass and spin interval, the examples were selected at random. The contrast between high- and low-spin systems is visible in the edge-on velocity maps over the full $3\,R_{1/2,\star}$ aperture. Closer to $R_{1/2,\star}$, particularly for the lowest-mass galaxies, the smaller number of stellar particles makes the velocity structure less clearly sampled. This behaviour is consistent with our use of the $3\,R_{1/2,\star}$ aperture adopted by \citet{Lagos2017} and later applied to TNG100 by \citet{Pallero2025}. The maps are constructed directly from stellar particles and should not be regarded as mock integral-field observations; reproducing the spatial sampling, seeing, and signal-to-noise of an IFS datacube would require substantially higher mass and spatial resolution.

For this test, we also estimate the stellar ellipticity following the
procedure of \citet{Pallero2025}.
The stellar inertia tensor is diagonalised and its eigenvalues are used
to derive the principal axes of the galaxy. The ellipticity is defined as
\begin{equation}
    \varepsilon = 1 - \frac{c}{a},
\end{equation}
where $a$ and $c$ are the major and minor axes, respectively.

\begin{figure}
\centering
\includegraphics{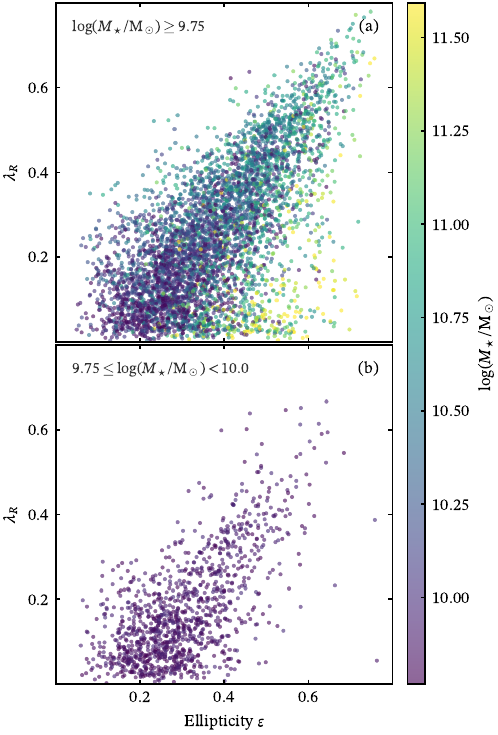}
\caption{Spin parameter $\lambda_R$ as a function of ellipticity $\varepsilon$ for simulated satellite galaxies at $z=0$, shown in the same plane used for the SAMI cluster galaxies by \citet{Brough2017}. Panel (a) shows the full TNG-Cluster+TNG300 satellite sample above the stellar-mass limit $\log(M_\star/\mathrm{M_\odot}) \geq 9.75$, while panel (b) is restricted to the lowest-mass bin, $9.75 \leq \log(M_\star/\mathrm{M_\odot}) < 10.0$. Points are colour-coded by stellar mass using a common scale spanning the 2nd--98th percentiles. For visual clarity, 15 per cent of the galaxies in each selection are shown. The simulated $\lambda_R$ values are measured within three stellar half-mass radii (Section~\ref{spin parameter simulation}).}
\label{fig:figC2}
\end{figure}

Figure~\ref{fig:figC2} shows the simulated galaxies in the $\lambda_R$--$\varepsilon$ plane used for the SAMI cluster sample by \citet{Brough2017}. Panel~(a) contains the full simulated sample, while panel~(b) is restricted to the lowest-mass bin, $9.75 \leq \log(M_\star/\mathrm{M_\odot}) < 10.0$. In both cases, rounder galaxies tend to occupy lower $\lambda_R$, whereas more flattened systems extend to higher values. The lowest-mass galaxies follow the same broad trend as the full sample, despite being resolved with the fewest stellar particles. The comparison is intended to be qualitative. Because the simulated $\lambda_R$ values are measured within $3\,R_{1/2,\star}$ rather than within one effective radius, their absolute values are not directly equivalent to those measured from SAMI. Figure~\ref{fig:figC2} is used only as a consistency check on the behaviour of the simulated spin proxy.

\begin{figure*}
\centering
\rotatebox{90}{%
  \includegraphics[height=\textwidth,width=0.90\textheight,keepaspectratio]{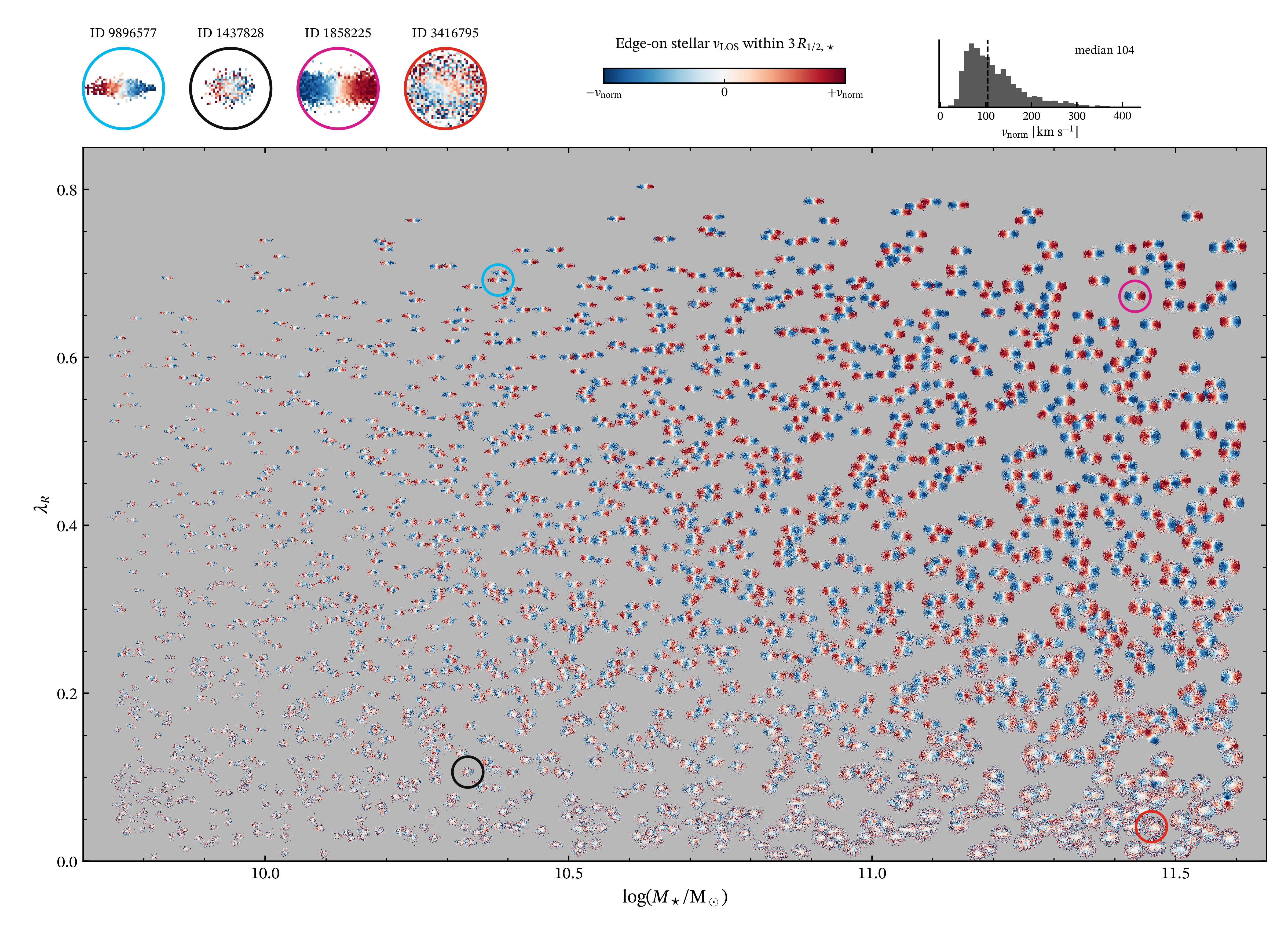}%
}
\caption{Edge-on stellar line-of-sight velocity maps of TNG-Cluster galaxies at $z=0$, placed according to their position in the $\lambda_R$--$\log(M_\star)$ plane. Each vignette shows stellar particles within $3\,R_{1/2,\star}$, the aperture used to measure $\lambda_R$ (Section~\ref{spin parameter simulation}). The $v_{\rm LOS}$ scale is normalised independently for each galaxy to $\pm v_{\rm norm}$; the distribution of $v_{\rm norm}$ is shown in the upper-right histogram, with the median marked by the dashed line. Where vignettes overlap, their plotting order is randomised. Four galaxies spanning the plane are highlighted by coloured circles and shown at larger scale in the upper strip, where they are labelled by subhalo ID. The displayed galaxies are a small subsample of the full population of $\sim 30\,000$ satellites, selected to provide approximately uniform coverage of the $\lambda_R$--$M_\star$ plane. Their vignette density therefore does not represent the underlying satellite distribution.}
\label{fig:figC3}
\end{figure*}

\begin{figure*}
\centering
\includegraphics{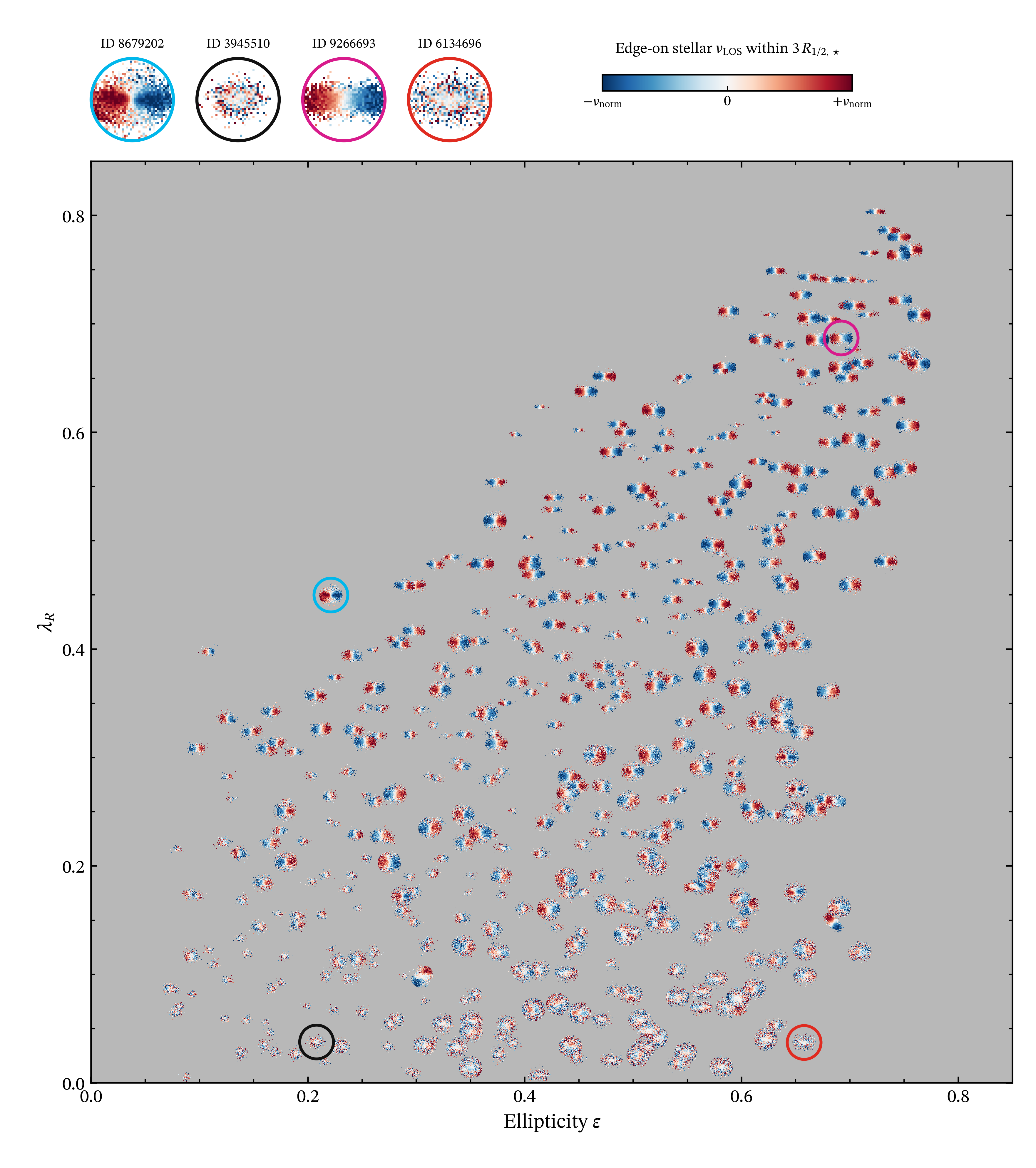}
\caption{Same as Fig.~\ref{fig:figC3}, but with galaxies placed in the $\lambda_R$--$\varepsilon$ plane shown in Fig.~\ref{fig:figC2}. The four highlighted galaxies are again shown enlarged in the upper strip. To reduce overlap along the main locus of the distribution, the displayed sample was thinned by retaining at most two galaxies per cell of a regular grid in this plane. The vignette density therefore does not represent the underlying satellite distribution.
\label{fig:figC4}}
\end{figure*}

Figures~\ref{fig:figC3} and~\ref{fig:figC4} provide a complementary view of this behaviour across the sample. The edge-on stellar $v_{\rm LOS}$ field of each selected galaxy is placed at its location in the $\lambda_R$--$\log(M_\star)$ and $\lambda_R$--$\varepsilon$ planes, allowing the velocity structure associated with different $\lambda_R$ values to be inspected directly. Coherent velocity gradients are common at high $\lambda_R$, while systems at low $\lambda_R$ generally show weaker or less ordered patterns. This behaviour is seen across the stellar-mass range rather than appearing only above a particular mass. The lowest-mass maps are visibly more sparsely sampled, as expected from their smaller stellar-particle counts, which further motivates treating the simulated $\lambda_R$ values as proxies for rotational support rather than as detailed IFS-like measurements.

\bsp
\label{lastpage}

\end{document}